%% file: ConsInt.tex
\documentclass[12pt,a4paper]{article}
\usepackage[authoryear]{natbib}
\usepackage{amsmath}
\usepackage{graphicx}
\usepackage{pdflscape}
\newcommand\auslassen[1]{}

\usepackage{multirow,booktabs,caption}
\DeclareCaptionLabelSeparator*{spaced}{\\[0.1ex]}
\usepackage[breaklinks=true,allcolors=black,colorlinks=true]{hyperref}

\begin{document}

\input{TitlePage}

\section{Introduction}
\input{Intro}

\section{Constraint Interaction}
\label{sec:CI}
\input{ConstIntro}
\subsection{Constraint Interaction: A First Example}
\label{subsec:ExampleTwo}
\input{ConstraintKline}

\subsection{Constraint Interaction: A Second Example}
\label{subsec:ExampleFour}
\input{Constraint}

\section{Interpreting Estimated Model Parameters under Different Scaling Methods}
\label{sec:Interpret}
\input{Interpretation}

\section{Explaining Constraint Interaction}
\label{sec:Explain}
\input{Explanation}

\section{Conclusion}
\label{sec:Conc}
\input{Conclusion}

\subsection*{Acknowledgements}
The authors are grateful for valuable remarks from Sandra Baar, Martin Becker, and Mireille Soliman.



\addcontentsline{toc}{section}{References}
\bibliographystyle{apalike}
\bibliography{export}

\begin{appendix}
\renewcommand{\thesection}{Appendix~\arabic{section}}
\section{}
\label{app:Formulas}
\input{Appendix}

\clearpage
\section{}
\label{app:Figures}
\input{AdditionalFigures}

\end{appendix}

\end{document}

%% file: TitlePage.tex
\author{
Stefan Kl\"o{\ss}ner\\
Statistics and Econometrics \\ 
Saarland University\\
\and
Eric Klopp\thanks{Corresponding author: 
Eric Klopp, Saarland University, Department of Education, Bldg.\ A 4.2, 66123 Saarbr\"ucken, Germany, email: e.klopp@mx.uni-saarland.de.} \\
Department of Education\\ 
Saarland University\\
}
\title{Explaining Constraint Interaction: 
How to Interpret Estimated Model Parameters under Alternative Scaling Methods}
\date{\today}
\maketitle
\begin{abstract}
In this paper, we explain the reasons behind constraint interaction, which is the phenomenon that the results of
testing equality constraints may depend heavily on the scaling method used. 
We find that the scaling methods interfere with the testing procedures 
because scaling methods determine which transformations of population quantities model parameters actually estimate. 
We therefore also develop rules on how to correctly interpret estimates of 
model parameters under alternative scaling methods.\\

{\bf Keywords}: constraint interaction, equality constraints, interpretation of parameter estimates\\
\end{abstract}

\clearpage

%% file: Intro.tex
In structural equation modeling, it is well known that latent variables must be given a scale for 
the model to have a chance to be identified. In the literature, three different 
methods for scaling latent factors are discussed: 
setting the loading of one indicator per factor to unity 
(fixed marker method, also called unit loading identification, ULI), 
setting the latent variables' variances to unity
(fixed factor method, also called unit variance identification, UVI),
and imposing the restriction that the average loading of every factor's indicators equals unity
(effects coding method, \citealp{LiCaSl}). 
In applications, it may happen that the result of 
testing hypotheses about the model parameters depends on the scaling method that is employed to carry out the model estimations:
for instance, a statistical test of the null hypothesis whether two indicators load equally strongly 
on their respective factors may be accepted when the fixed marker method is used, but the same 
hypothesis may be rejected when the fixed factor method is used.
This phenomenon, which has been introduced to the literature by \cite{Steiger}, is called constraint interaction. 

The reasons underlying constraint interaction have not yet been fully explored in the literature. 
In order to close this gap, 
we revisit constraint interaction in the context of CFA models and
elaborate on the causes of this phenomenon.\footnote{Constraint 
interaction may also appear in more general structural equation models. Studying these, however,
is beyond the scope of this paper.} We find that 
constraint interaction is intimately linked to the interpretation of parameter estimates under
alternative scaling methods, because the latter determine which population quantity model parameters actually estimate. 
We therefore develop rules that show how to correctly interpret estimates of factor loadings and other model parameters 
under alternative scaling methods. These rules are not only important for understanding constraint interaction,
they also help practitioners to interpret estimated models correctly and to avoid pitfalls when 
drawing conclusions from estimated model parameters.

Key to understanding constraint interaction is the fact that the quantities that model parameters 
actually estimate depend on the scaling method used for achieving model identification. For instance, 
when testing equality of two loading parameters, a corresponding test procedure using the fixed marker method
for scaling the factors will actually test whether, in the population, the ratios of the corresponding loadings over the
marker variables' loadings coincide \citep{RaMaLi}. In contrast, when using the fixed factor method, trying to 
test for identical loadings will in fact lead to testing whether the products of loading and factor standard deviation
are identical in the population. Therefore, constraint interaction occurs because different hypotheses are tested empirically, although 
this fact does not become obvious to the researcher. For this reason, whenever researchers encounter constraint interaction in practice, 
they should 
take great care in making sure how or whether at all the originally intended hypothesis may be tested empirically.

The paper is structured as follows: 
in Section~\ref{sec:CI}, we showcase the phenomenon of constraint interaction with the help of an example given by \cite{Kline}
and an example from a longitudinal context. 
Section~\ref{sec:Interpret} studies which population quantities are estimated by model parameters,
depending on the scaling method employed.
In Section~\ref{sec:Explain}, we investigate the reasons behind constraint interaction, 
while Section~\ref{sec:Conc} concludes.

%% file: ConstIntro.tex
In this section, the problem of constraint interaction will be exemplified
in the context of confirmatory factor analysis (CFA).
We first discuss a simple example on constraint interaction given in \citet[p.~336f]{Kline}. 
This example provides an illustration of constraint interaction in the context of a two-factor model 
in which the hypothesis of the equality of two factor loadings is tested using various scaling methods. 
As
the initial example is rather simplistic, we subsequently consider a longitudinal one-factor model with a larger number of manifest indicators. 

%% file: ConstraintKline.tex
\citet[p.~336f]{Kline} considers a two-factor CFA model where the factors $A$ and $B$ have two indicators each:
the indicators of factor $A$ are $X_1$ and $X_2$ and the indicators of factor $B$ are 
$X_3$ and $X_4$, as displayed by Figure~\ref{fig:ParamsQuantTwo}. 
With this example, we introduce the notation used throughout this paper: 
expressions using Latin letters refer to population quantities, 
while expressions using Greek letters refer to model parameters which are estimated using 
different scaling methods. 
Thus, $A \rightarrow X_i$, $\operatorname{Var}(E_i)$, $\operatorname{Var}(A)$, $\operatorname{Var}(B)$, and $\operatorname{Cov}(A,B)$ refer to the corresponding population quantities, 
while $\lambda_i$, $\Theta_{ii}$, $\Phi_{AA}$, $\Phi_{BB}$, and $\Phi_{AB}$ refer to model parameters ($i=1,\ldots,4)$.

\auslassen{
\begin{figure}[ht]
\subfigure[Population quantities]{\includegraphics[width=0.495\textwidth]{First_Example/Unrestricted_Quantities}}
\subfigure[Model parameters]{\includegraphics[width=0.495\textwidth]{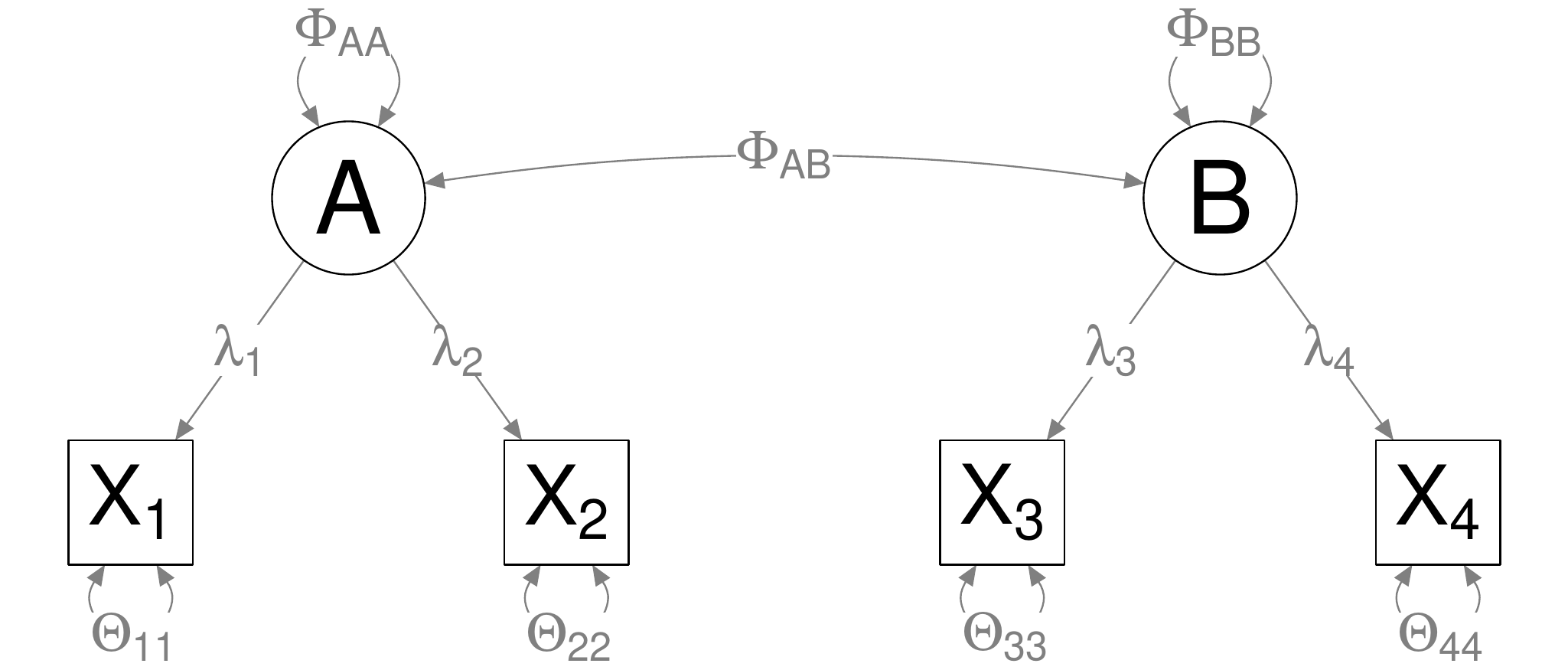}}
\caption{Population quantities and model parameters for the example by \cite{Kline}.} 
\label{fig:ParamsQuantTwo}
\end{figure}
}

\begin{figure}[ht]
\includegraphics[width=0.99\textwidth]{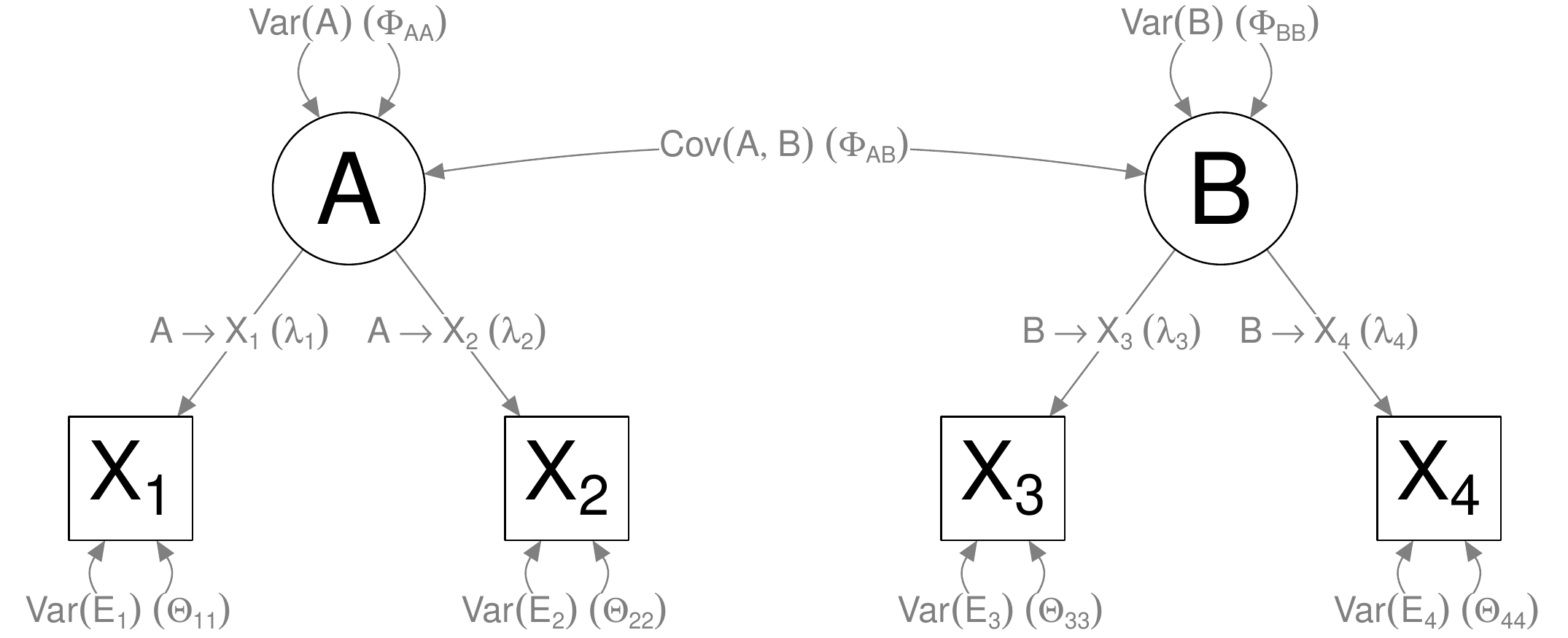}
\caption{Population quantities and their respective model parameters (in parentheses) for the example by \cite{Kline}.} 
\label{fig:ParamsQuantTwo}
\end{figure}

The question under scrutiny is whether, in the population, 
'the unstandardized pattern coefficients of $X_2$ and $X_4$ are equal' \citep[p. 336]{Kline}. We are thus interested
in the hypotheses
\begin{equation}
\label{eq:H0_Two}
H_0: A \rightarrow X_2 \,\, = \,\, B\rightarrow X_4\quad\mbox{ vs. } \quad
H_1: A \rightarrow X_2 \,\, \neq \,\, B\rightarrow X_4.
\end{equation}

It is essential that the hypotheses in equation~\eqref{eq:H0_Two}
are formulated in terms of the population quantities $A\to X_2$ and $B\to X_4$, 
as these are the fundamental, unobservable quantities about which we want to draw conclusions,
while $\lambda_2$ and $\lambda_4$ merely denote model parameters whose values are estimated from the data at hand. 

\citet[p.~336f]{Kline} further specifies that from $N=200$ data points, 
the covariance matrix of the indicators $X_1,X_2,X_3,X_4$ was estimated as\footnote{In all estimations below, 
$S$ is interpreted as a maximum likelihood estimate of the covariance matrix, 
i.e.\ as having been calculated using the number of data points in the denominator of the corresponding formula.
}
\begin{equation}
\label{eq:SigmaTwo}
S=
\begin{pmatrix}
25.00 & & & \\
7.20 & 9.00 & & \\
3.20 & 2.00 & 4.00 & \\
2.00 & 1.25 & 1.20 & 4.00\\
\end{pmatrix}.
\end{equation}

For testing the hypotheses defined in~\eqref{eq:H0_Two}, two nested CFA models are estimated: 
in the first step, the original model is estimated imposing only a scaling condition, but 
without imposing any restriction related with
the null hypothesis $H_0$ in~\eqref{eq:H0_Two} ('unrestricted estimation'). 
In the second step, the restriction given by the null hypothesis $H_0$ in~\eqref{eq:H0_Two} 
is additionally imposed ('restricted estimation'), 
by adding the condition that $\lambda_2$ must equal $\lambda_4$, which is the direct translation 
of the null hypothesis $A\to X_2 = B\to X_4$ in terms of model parameters. 
Thus, when using the fixed marker method, $\lambda_1$ and $\lambda_3$ are fixed to unity under both
the unrestricted and restricted estimation, while the latter also comprises the condition that $\lambda_2=\lambda_4$.
In contrast, when using the fixed factor method, $\Phi_{AA}$ and $\Phi_{BB}$ are fixed to unity under
both the unrestricted and restricted estimation, and the latter again also comprises the condition that $\lambda_2=\lambda_4$.
Finally, for the effects coding method, both unrestricted and restricted model estimation are carried out under 
the condition that both $\frac{\lambda_1+\lambda_2}{2}$ and $\frac{\lambda_3+\lambda_4}{2}$ are equal to unity,
and the restricted estimation additionally features the condition that $\lambda_2=\lambda_4$.

Table~\ref{tab:ResultsTwo} contains the fit indices for the unrestricted and restricted estimations:\footnote{Estimations were carried out 
using the freely available statistical software R, version 3.3.2, in combination with package lavaan, version 0.5-22, see \cite{R} and \cite{lavaan}.}
all unrestricted models fit the data equally well, namely perfectly. 
In contrast, the results of the restricted estimations show a different pattern: 
the restricted models under the fixed marker and effects coding scaling indicate perfect model fit, 
whereas the restricted model under fixed factor scaling indicates a very bad model fit, 
as can be seen from the significant $\chi^2$-statistics as well as from the values of CFI, RMSEA, and SRMR.


In general, hypotheses about the equality of parameters can be tested by means of $\chi^2$-difference tests as well as by 
investigating how much fit indices deteriorate when imposing the equality constraint.\footnote{For ease of exposition, 
we mainly focus on the $\chi^2$-difference-statistics and the corresponding $p$-values. 
This does not cause any loss of generality, because constraint interaction emerges in exactly the same way when changes of fit indices are used
for testing equality constraints.}
Following this procedure, the $\chi^2$-difference tests and fit index differences under the fixed marker and effects coding scaling provide empirical evidence in support of the 
null hypothesis $H_0$ given in~\eqref{eq:H0_Two}. Thus, one would conclude that the population loadings $A\to X_2$ and $B\to X_4$ coincide.
In contrast, the $\chi^2$-difference test and the fit index differences under the fixed factor scaling provide empirical evidence against the null hypothesis $H_0$ given in~\eqref{eq:H0_Two}. 
Consequently, one would conclude that $H_1$ in~\eqref{eq:H0_Two} is better in accordance with the given data and that $A\to X_2$ is not equal to $B\to X_4$. 
The example at hand thus is a prototypical example of constraint interaction, as introduced by \cite{Steiger}: the result of investigating 
equality constraints depends on which method is used for scaling the factors.

\input{Table_TwoIndicators}

%% file: Table_TwoIndicators.tex
\begin{table}[ht]
\centering
\caption{Fit measures of unrestricted and restricted estimation depending on alternative scaling \\ methods and the respective fit measures' differences for the first example.} 
\label{tab:ResultsTwo}
\begin{tabular}{llrrr}
  \toprule
Model & Fit Measure & Marker & Factor & Effects \\ 
  \midrule
\multirow{6}{*}{Unrestricted} & $\chi^2$ & 0.00000 & 0.00000 & 0.00000 \\ 
   & $df$ & 1 & 1 & 1 \\ 
   & $p$ & 1.00000 & 1.00000 & 1.00000 \\ 
   & CFI & 1.00000 & 1.00000 & 1.00000 \\ 
   & RMSEA & 0.00000 & 0.00000 & 0.00000 \\ 
   & SRMR & 0.00000 & 0.00000 & 0.00000 \\ 
   \midrule
\multirow{6}{*}{Restricted} & $\chi^2$ & 0.00000 & 14.08728 & 0.00000 \\ 
   & $df$ & 2 & 2 & 2 \\ 
   & $p$ & 1.00000 & 0.00087 & 1.00000 \\ 
   & CFI & 1.00000 & 0.87958 & 1.00000 \\ 
   & RMSEA & 0.00000 & 0.17383 & 0.00000 \\ 
   & SRMR & 0.00000 & 0.09790 & 0.00000 \\ 
   \midrule
\multirow{6}{*}{Difference} & $\Delta \chi^2$ & 0.00000 & 14.08728 & 0.00000 \\ 
   & $\Delta df$ & 1 & 1 & 1 \\ 
   & $p$ & 1.00000 & 0.00017 & 1.00000 \\ 
   & $\Delta$CFI & 0.00000 & -0.12042 & 0.00000 \\ 
   & $\Delta$RMSEA & 0.00000 & 0.17383 & 0.00000 \\ 
   & $\Delta$SRMR & 0.00000 & 0.09790 & 0.00000 \\ 
   \bottomrule
\end{tabular}
\end{table}

%% file: Constraint.tex
As a second example for constraint interaction, 
we will now 
discuss a more complex model containing 
a factor with four indicators, which in the context of longitudinal modeling
is observed twice over time. At time $1$, factor $A$ is denoted by $A_1$ and measured by four indicators, 
$X_{11}$, $X_{21}$, $X_{31}$, and $X_{41}$, 
while $A$'s unobservable value at time $2$ is denoted by
$A_2$ and analogously measured by four indicators,
$X_{12}$, $X_{22}$, $X_{32}$, and $X_{42}$.
As is common in longitudinal modeling, 
errors belonging to repeated measurements 
are allowed to correlate, i.e.\
$E_{11}$ may correlate with $E_{12}$, $E_{21}$ with $E_{22}$, $E_{31}$ with $E_{32}$, 
and $E_{41}$ with $E_{42}$. 
The corresponding model is shown graphically in Figure~\ref{fig:ParamsQuantFour}.
\auslassen{
\begin{figure}[ht]
\centering
\includegraphics[width=0.9\textwidth]{Unrestricted_QuantitiesII}
\caption{Longitudinal CFA model with population quantities.}
\label{fig:ParamsQuantFour}
\end{figure}
}%
\begin{figure}[ht]
\centering
\includegraphics[width=0.99\textwidth]{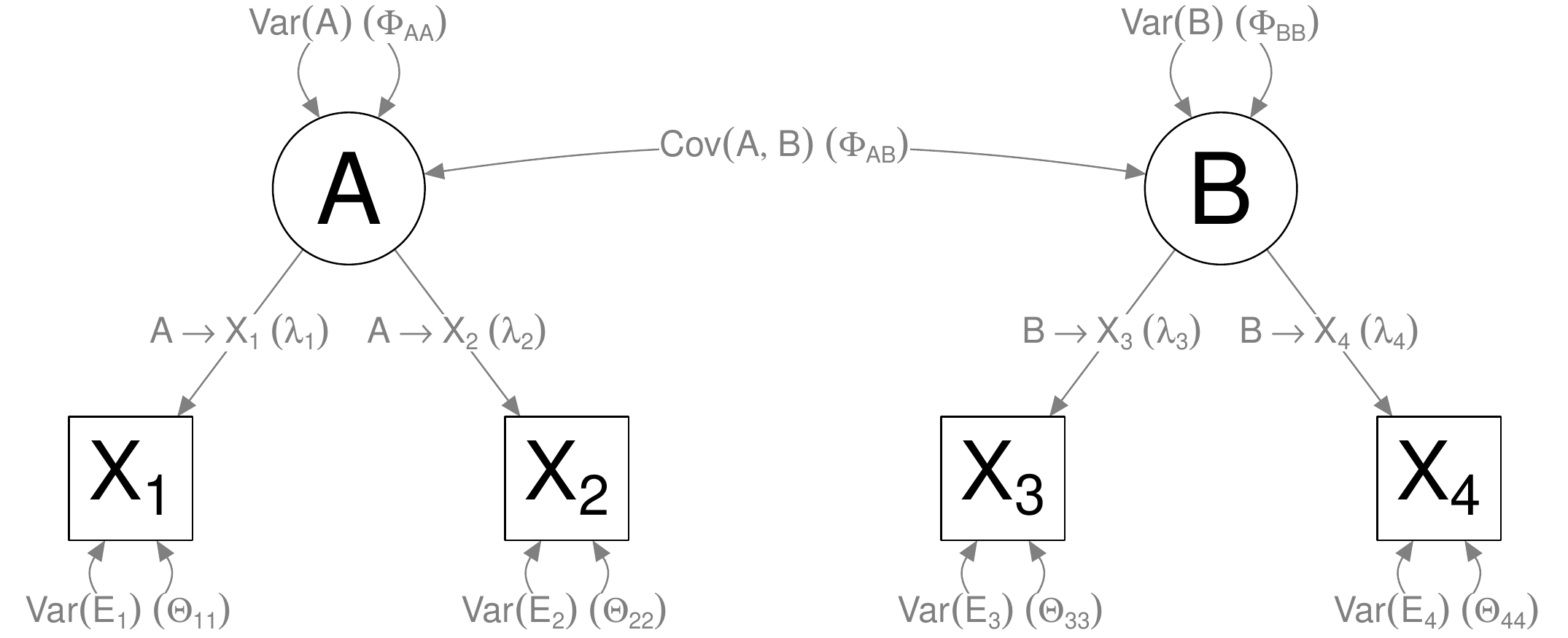}
\caption{Longitudinal CFA model with population quantities and their respective model parameters (in parentheses).}
\label{fig:ParamsQuantFour}
\end{figure}
The question under scrutiny is whether the factor loadings of the second indicators, $X_{21}$ and $X_{22}$, are equal
in the population, i.e.\ we consider the hypotheses
\begin{equation}
\label{eq:H0_Four}
H_0: A_1 \rightarrow X_{21} \,\, = \,\, A_2\rightarrow X_{22}\quad\mbox{ vs. }\quad
H_1: A_1 \rightarrow X_{21} \,\, \neq \,\, A_2\rightarrow X_{22}.
\end{equation}

From $N=150$ data points, the covariance matrix of the indicators 
$X_{11}$, $X_{21}$, $X_{31}$, $X_{41}$, 
$X_{12}$, $X_{22}$, $X_{32}$, $X_{42}$, 
was estimated as
\begin{equation}
\label{eq:SigmaFour}
S=
\begin{pmatrix}
3.640 & & & & & & & \\ 
  3.200 & 17.000 & & & & & & \\ 
  2.560 & 12.800 & 14.240 & & & & & \\ 
  1.600 & 8.000 & 6.400 & 6.000 & & & & \\ 
  1.160 & 4.800 & 3.840 & 2.400 & 27.000 & & & \\ 
  0.960 & 5.300 & 3.840 & 2.400 & 25.000 & 32.000 & & \\ 
  0.768 & 3.840 & 3.322 & 1.920 & 20.000 & 20.000 & 17.000 & \\ 
  0.384 & 1.920 & 1.536 & 1.460 & 10.000 & 10.000 & 8.000 & 12.000 \\ 
\end{pmatrix}.
\end{equation}

The procedure for testing~\eqref{eq:H0_Four} is the same as in the first example above:
estimating the unrestricted and restricted version of the model under different scaling methods and analyzing
the fit's worsening due to imposing the restriction, Figure~\ref{fig:ParamsQuantFour} displays which parameters are used
for these estimations.
With respect to available scaling methods, 
there are now at least three versions of the fixed marker method,
because each factor has four indicators:
we can use the first, third, and fourth indicators as marker variables.
Thus, when using the first indicators as marker variables, 
the restrictions $\lambda_{11}=1$ and $\lambda_{12}=1$ 
are used, while $\lambda_{31}$ and $\lambda_{32}$ ($\lambda_{41}$ and $\lambda_{42}$) are set to unity 
when the third (fourth) indicators take the role of marker variables.\footnote{Of course, one may also 
use for instance the first indicator, $X_{11}$, for scaling the first factor, $A_1$, while using 
the third indicator, $X_{32}$, for scaling the second factor, $A_2$. Such a 'mixed' choice of marker variables,
however, is rarely used in practice. We thus refrain from including these variants, although it would be
perfectly possible to do so.}
The fixed factor method is characterized by imposing the restrictions $\Phi_{11}=1$ and $\Phi_{22}=1$,
while the effects coding method constrains $\frac{\lambda_{11}+\hdots+\lambda_{41}}{4}$ and 
$\frac{\lambda_{12}+\hdots+\lambda_{42}}{4}$
to unity.
For all scaling methods, the restricted model for testing the null hypothesis
$A_1 \rightarrow X_{21} \,\, \stackrel{?}{=} \,\, A_2\rightarrow X_{22}$
is derived from the unrestricted one by adding the constraint $\lambda_{21}=\lambda_{22}$.

\auslassen{
\begin{figure}[ht]
\centering
\includegraphics[width=0.9\textwidth]{Second_Example/Unrestricted_Params}
\caption{Longitudinal CFA model with model parameters.}
\label{fig:ParamsFour}
\end{figure}
}

Summaries of the estimation results for the unrestricted models are provided in the upper part of Table~\ref{tab:ResultsFour}:
for all five alternative scalings, the $\chi^2$-statistics as well as CFI, RMSEA, and SRMR indicate a perfect model fit. Thus, the results of the unrestricted model with respect to fit
measures do not depend on the scaling method.
\input{Table_VersionC}

The results of estimating the restricted model are displayed in the middle part of 
Table~\ref{tab:ResultsFour}:
the results of the restricted estimation are inconsistent and depend on the used scaling method. 
With respect to the fixed marker scaling, using the first indicators as markers indicates a very bad model fit, 
while using the third indicators as markers indicates perfect model fit. 
Using the fourth indicators as markers, the restricted model shows a nearly perfect model fit. 
Under the fixed factor scaling, the restricted model also displays a very good model fit. 
In case of the effects coding scaling, the evaluation of the model fit using the $\chi^2$-statistic depends on the nominal significance level: 
using the 5\%-level, one would reject the restricted model, whereas one would accept it when using a 1\%-significance level. 

Consequently, the results of the nested-model comparisons also differ, revealing that constraint interaction occurs in this example, too:
there is extremely strong empirical evidence against the null hypothesis $H_0$ of loading equality given in equation~\eqref{eq:H0_Four}
when the first indicators are used as marker variables, while the data are perfectly in line with this hypothesis when the third indicators are used as markers.
Between those two extremes, using the fourth indicators as markers produces a $p$ value of $0.098$ and leads to accepting the null hypothesis at 
common significance levels of $5\%$ or $1\%$, while the decision about the hypotheses in case of fixed factor scaling depends on the nominal significance level, due to a $p$ value of $0.01814$. 
Finally, if the effects coding scaling is used, there is evidence against the null hypothesis, albeit not as strong as when the first indicators are used as markers.

Overall, the example thus proves that constraint interaction may occur also in longitudinal studies and that it is not restricted to simplistic models like the one in the first example. 
To the contrary, constraint interaction is quite likely to occur in many applications of CFA (and more general in structural equation models) in practice.\footnote{We will elaborate on the reasons for the likely occurrence of constraint interaction in practice in section~\ref{sec:Explain} below.}

%% file: Table_VersionC.tex
\begin{table}[ht]
\centering
\caption{Fit measures of unrestricted and restricted estimation depending on alternative scaling \\ methods and the respective fit measures' differences for the second example.} 
\label{tab:ResultsFour}
\begin{tabular}{llrrrrr}
  \toprule
Model & Fit Measure & Marker 1 & Marker 3 & Marker 4 & Factor & Effects \\ 
  \midrule
\multirow{6}{*}{Unrestricted} & $\chi^2$ & 0.00000 & 0.00000 & 0.00000 & 0.00000 & 0.00000 \\ 
   & $df$ & 15 & 15 & 15 & 15 & 15 \\ 
   & $p$ & 1.00000 & 1.00000 & 1.00000 & 1.00000 & 1.00000 \\ 
   & CFI & 1.00000 & 1.00000 & 1.00000 & 1.00000 & 1.00000 \\ 
   & RMSEA & 0.00000 & 0.00000 & 0.00000 & 0.00000 & 0.00000 \\ 
   & SRMR & 0.00000 & 0.00000 & 0.00000 & 0.00000 & 0.00000 \\ 
   \midrule
\multirow{6}{*}{Restricted} & $\chi^2$ & 145.54167 & 0.00000 & 2.72338 & 5.58214 & 26.66053 \\ 
   & $df$ & 16 & 16 & 16 & 16 & 16 \\ 
   & $p$ & 0.00000 & 1.00000 & 0.99991 & 0.99201 & 0.04541 \\ 
   & CFI & 0.85774 & 1.00000 & 1.00000 & 1.00000 & 0.98829 \\ 
   & RMSEA & 0.23233 & 0.00000 & 0.00000 & 0.00000 & 0.06665 \\ 
   & SRMR & 0.24422 & 0.00000 & 0.03242 & 0.09026 & 0.07406 \\ 
   \midrule
\multirow{6}{*}{Difference} & $\Delta \chi^2$ & 145.54167 & 0.00000 & 2.72338 & 5.58214 & 26.66053 \\ 
   & $\Delta df$ & 1 & 1 & 1 & 1 & 1 \\ 
   & $p$ & 0.00000 & 1.00000 & 0.09889 & 0.01814 & 0.00000 \\ 
   & $\Delta$CFI & -0.14226 & 0.00000 & 0.00000 & 0.00000 & -0.01171 \\ 
   & $\Delta$RMSEA & 0.23233 & 0.00000 & 0.00000 & 0.00000 & 0.06665 \\ 
   & $\Delta$SRMR & 0.24422 & 0.00000 & 0.03242 & 0.09026 & 0.07406 \\ 
   \bottomrule
\end{tabular}
\end{table}

%% file: Interpretation.tex
%

In order to lay the ground for explaining the reasons behind constraint interaction, this section is devoted
to studying how different scaling methods affect which population quantities (or transformations thereof)
model parameters actually estimate.
To this end, we first reconsider the 
initial example of a two-factor CFA
with two indicators per factor, as this example is the less complex one. Building on the results
obtained for the simpler case, we then proceed by studying the much more general example
of four indicators per factor in a longitudinal context.

\subsection{Reconsidering the first example}

Table~\ref{tab:ParametersTwo} shows, for the unrestricted model, the estimated parameter values of
loadings, factor (co-)variances, and residual variances.
Both estimated loadings and (co-)va\-ri\-an\-ces of the factors change when the scaling method is altered, 
while the estimated residual variances of the indicators are invariant 
to the scaling method. 
Thus, only the residual variances can be interpreted 'as is', i.e.\ without taking into account
which scaling method has been applied.
\input{TableResults_TwoIndicators}

With respect to loading parameters, \cite{Newsom} shows how the values obtained under the fixed marker scaling are related to the corresponding values
under fixed factor scaling: for instance, multiplying a loading's value calculated using the fixed factor method by the square root of the variance parameter's value 
for the fixed factor scaling produces the loading's value obtained under the fixed marker method \citep[Formula 1.5]{Newsom}. Similarly,
the estimated loading under the fixed marker method can be obtained as the ratio of the values of the loading of interest and the loading of
the referent indicator, where the latter loadings are calculated using the fixed factor scaling \citep[Formula 1.5]{Newsom}.
Table~\ref{tab:Identified} presents the estimated values for these (and other) transformations of model parameters, 
for all three scaling methods:\footnote{Table~\ref{tab:Identified} and other lengthy tables have been relegated to \ref{app:Figures}.}
it becomes evident that the method of scaling does not impact the values estimated for these transformations of parameters, 
i.e.\ the corresponding values are invariant of the scaling method employed. As a result, these transformations of model parameters
can be interpreted 'as is': for instance, the ratio of two loading parameters measures the corresponding ratio of population quantities,
e.g., $\frac{\lambda_2}{\lambda_1}$ estimates $\frac{A\to X_2}{A\to X_1}$. Table~\ref{tab:Identified} can therefore be used to derive
which transformations of population quantities are actually estimated by model parameters, depending on the respective scaling method.


With regard to the fixed marker scaling, Table~\ref{tab:Identified} reveals that the ratios
$\frac{\lambda_2}{\lambda_1}$ and $\frac{\lambda_4}{\lambda_3}$ do not depend on the scaling method.
Therefore, they always estimate their corresponding population counterpart, 
i.e.\ $\frac{\lambda_2}{\lambda_1}$ and $\frac{\lambda_4}{\lambda_3}$
always estimate $\frac{A\rightarrow X_2}{A\rightarrow X_1}$ and $\frac{B\rightarrow X_4}{B\rightarrow X_3}$, respectively, 
regardless of which scaling method is used. 
As a consequence, 
under the fixed marker method, $\lambda_2$ and $\lambda_4$ are estimates of these quantities,
due to fixing $\lambda_1$ and $\lambda_3$ to unity.
Put differently, using $X_1$ as marker variable leads to 
$\lambda_2=\frac{\lambda_2}{1}=\frac{\lambda_2}{\lambda_1}$ 
estimating the ratio of $X_2$'s and $X_1$'s loading on factor $A$,
and choosing $X_3$ for scaling factor $B$ entails that
$\lambda_4=\frac{\lambda_4}{1}=\frac{\lambda_4}{\lambda_3}$ 
estimates the ratio of $X_4$'s and $X_3$'s loading on factor $B$. The general rule for estimated factor loadings under the fixed marker scaling is that they estimate the ratio between an indicator's loading and that of the marker indicator, as also observed by \citet{RaMaLi}.
Following this rule, the correct interpretation of $\lambda_2=0.625$ under the fixed marker method is
\emph{not} that 
"$X_2$'s factor loading on $A$ is $0.625$", but that
"$X_2$ loads on $A$ $0.625$ as much as does $X_1$".

Concerning the fixed factor scaling, Table~\ref{tab:Identified} shows that the terms $\Phi_{AA}\lambda_1^2$, $\Phi_{BB}\lambda_3^2$, and 
$\Phi_{AB}\lambda_1\lambda_3$ do not depend on the scaling method. Therefore, they are
estimates of $\operatorname{Var}(A)\cdot\left(A\rightarrow X_1\right)^2$,
$\operatorname{Var}(B)\cdot\left(B\rightarrow X_3\right)^2$,
and $\operatorname{Cov}(A,B)\cdot\left(A\rightarrow X_1\right)\cdot\left(B\rightarrow X_3\right)$, respectively.
Furthermore, these are exactly the quantities that $\Phi_{AA}$, $\Phi_{BB}$, and $\Phi_{AB}$ estimate when 
$X_1$ and $X_3$ are used as marker variables for $A$ and $B$, as then $\lambda_1=1$ and $\lambda_3=1$. 
Therefore, under the fixed marker method,
the correct interpretation of an estimated factor variance, e.g.\ 
$\Phi_{AA}=\Phi_{AA}\cdot 1^2=\Phi_{AA}\lambda_1^2=11.52$, 
is \emph{not} that "factor $A$'s variance is $11.52$", 
but that "the product of factor $A$'s variance and its marker variable's squared loading is $11.52$".
Correspondingly, the correct interpretation of $\Phi_{AB}=\Phi_{AB}\cdot 1\cdot 1=\Phi_{AB}\lambda_1\lambda_3=3.2$ 
under the fixed marker method is that 
"the product of the covariance of factors $A$ and $B$ and the loadings of their marker variables equals $3.2$".

Table~\ref{tab:Identified} also shows
that $\lambda_1\sqrt{\Phi_{AA}}$, $\lambda_2\sqrt{\Phi_{AA}}$, $\lambda_3\sqrt{\Phi_{BB}}$, and $\lambda_4\sqrt{\Phi_{BB}}$ 
estimate the quantities
$\left(A\rightarrow X_1\right)\cdot\sqrt{\operatorname{Var}(A)}$, 
$\left(A\rightarrow X_2\right)\cdot\sqrt{\operatorname{Var}(A)}$, 
$\left(B\rightarrow X_3\right)\cdot\sqrt{\operatorname{Var}(B)}$, 
and $\left(B\rightarrow X_4\right)\cdot\sqrt{\operatorname{Var}(B)}$,
irrespective of the scaling method.
Under the fixed factor method, therefore,
Tables~\ref{tab:ParametersTwo} and~\ref{tab:Identified} taken together show
that the model parameters for the factor loadings,
$\lambda_1=\lambda_1\sqrt{1}=\lambda_1\sqrt{\Phi_{AA}}$, $\lambda_2=\lambda_2\sqrt{1}=\lambda_2\sqrt{\Phi_{AA}}$, 
$\lambda_3=\lambda_3\sqrt{1}=\lambda_3\sqrt{\Phi_{BB}}$, and $\lambda_4=\lambda_4\sqrt{1}=\lambda_4\sqrt{\Phi_{BB}}$,
estimate the population quantities
$\left(A\rightarrow X_1\right)\cdot\sqrt{\operatorname{Var}(A)}$, 
$\left(A\rightarrow X_2\right)\cdot\sqrt{\operatorname{Var}(A)}$, 
$\left(B\rightarrow X_3\right)\cdot\sqrt{\operatorname{Var}(B)}$, 
and $\left(B\rightarrow X_4\right)\cdot\sqrt{\operatorname{Var}(B)}$.
Thus, the correct interpretation of $\lambda_1=\lambda_1\sqrt{1}=\lambda_1\sqrt{\Phi_{AA}}=3.39411$ under the fixed factor method is
\emph{not} that 
"$X_1$'s factor loading on $A$ is $3.39411$", but that 
"the product of $X_1$'s factor loading on $A$ and $A$'s standard deviation is $3.39411$".
Furthermore, Tables~\ref{tab:ParametersTwo} and~\ref{tab:Identified} 
show the well-known fact that $\Phi_{AB}$ estimates the correlation of $A$ and $B$ when the fixed factor method is used.

With regard to the effects coding scaling, 
Table~\ref{tab:Identified} shows
that
$\lambda_1$, $\lambda_2$, $\lambda_3$, and $\lambda_4$ estimate the quantities
$\frac{A\rightarrow X_1}{\frac{\left(A\rightarrow X_1\right)+\left(A\rightarrow X_2\right)}{2}}$,
$\frac{A\rightarrow X_2}{\frac{\left(A\rightarrow X_1\right)+\left(A\rightarrow X_2\right)}{2}}$,
$\frac{B\rightarrow X_3}{\frac{\left(B\rightarrow X_3\right)+\left(B\rightarrow X_4\right)}{2}}$,
and $\frac{B\rightarrow X_4}{\frac{\left(B\rightarrow X_3\right)+\left(B\rightarrow X_4\right)}{2}}$, respectively.
Thus, the correct interpretation of $\lambda_1=\frac{\lambda_1}{1}=\frac{\lambda_1}{\frac{\lambda_1+\lambda_2}{2}}=1.23077$ 
when using effects coding is
\emph{not} that 
"$X_1$'s factor loading on $A$ is $1.23077$", but
that "the ratio of $X_1$'s factor loading on $A$ and $A$'s indicators' average loading is $1.23077$" or, put differently,
that "$X_1$ loads $23.077$\% stronger on $A$ than $A$'s average indicator does".
Furthermore, Table~\ref{tab:Identified} reveals that, under effects coding,
$\Phi_{AA}$ and $\Phi_{BB}$ estimate the product of $A$ and $B$'s variance
and the squared average loading corresponding to that factor, respectively, while
$\Phi_{AB}$ estimates the covariance between factors $A$ and $B$ 
multiplied by these factors' average loadings.

For the reader's convenience, all the interpretations given above are
summarized in Table~\ref{tab:InterTwo} in the appendix.


\subsection{Reconsidering the second example}
We now turn our attention to interpreting the estimated parameters in the more complex longitudinal model featuring
one factor with four indicators measured twice over time. 
Again, we find that the estimated values for the loading parameters, 
$\lambda_{j1}, \lambda_{j2}$ ($j=1,\hdots,4$),
as well as those for the factors' (co-)variances, 
$\Phi_{11},\Phi_{22},\Phi_{12}$, strongly depend on the scaling method, 
see Table~\ref{tab:ParametersFour}.
In contrast, residual variances and covariances of indicators are invariant to changes of the scaling method, 
the corresponding parameters 
$\Theta_{j1,j1}$, $\Theta_{j2,j2}$, and $\Theta_{j1,j2}$ ($j=1,\hdots,4$)
thus estimate the corresponding population (co-)variances of the error terms, 
$\operatorname{Var}(E_{j1})$, $\operatorname{Var}(E_{j2})$, 
and $\operatorname{Cov}(E_{j1},E_{j2})$ ($j=1,\hdots,4$).

Combining the contents of Tables~\ref{tab:IdentifiedFourLambdas} and \ref{tab:IdentifiedFourLambdasII} with 
Table~\ref{tab:ParametersFour} shows which quantities the loading parameters estimate under different scaling methods:
when the first indicators are used as marker variables, 
$\lambda_{j1}=\frac{\lambda_{j1}}{1}=\frac{\lambda_{j1}}{\lambda_{11}}$ and 
$\lambda_{j2}=\frac{\lambda_{j2}}{1}=\frac{\lambda_{j2}}{\lambda_{12}}$ estimate the ratios
$\frac{A_1\to X_{j1}}{A_1\to X_{11}}$ and 
$\frac{A_2\to X_{j2}}{A_2\to X_{12}}$ ($j=1,\hdots,4$),
respectively. Similarly,
when the third (fourth) indicators are used as marker variables, $\lambda_{j1}$ and $\lambda_{j2}$ estimate 
$\frac{A_1\to X_{j1}}{A_1\to X_{31}}$ and $\frac{A_2\to X_{j2}}{A_2\to X_{32}}$
($\frac{A_1\to X_{j1}}{A_1\to X_{41}}$ and $\frac{A_2\to X_{j2}}{A_2\to X_{42}}$)
for $j=1,\hdots,4$. In general, thus, when the $i$-th indicators take the role of marker variables,
the $j$-th loading parameters estimate the ratios $\frac{A_1\to X_{j1}}{A_1\to X_{i1}}$ and 
$\frac{A_2\to X_{j2}}{A_2\to X_{i2}}$.
Therefore, for instance, the appropriate interpretation of $\lambda_{21}=5$ 
when the first indicator is used as marker variable is given by
"$X_{21}$ loads five times as strong on factor $A_1$ than $X_{11}$ does".

With respect to the fixed factor method, Tables~\ref{tab:IdentifiedFourLambdas} and \ref{tab:IdentifiedFourLambdasII} in combination with 
Table~\ref{tab:ParametersFour} reveal that the loading parameters $\lambda_{j1}=\lambda_{j1}\cdot 1=\lambda_{j1}\cdot\sqrt{\Phi_{11}}$ and 
$\lambda_{j2}=\lambda_{j2}\cdot 1=\lambda_{j2}\cdot\sqrt{\Phi_{22}}$ 
in this case estimate $(A_1\to X_{j1})\cdot\sqrt{\operatorname{Var}(A_1)}$ 
and $(A_2\to X_{j2})\cdot\sqrt{\operatorname{Var}(A_2)}$ ($j=1,\hdots,4$). 
The general rule for the interpretation of estimated loading parameters under the fixed factor method thus is that they estimate the 
product of the corresponding loading quantity in the population and the corresponding factor's standard deviation. 

For effects coding, Tables~\ref{tab:ParametersFour}-\ref{tab:IdentifiedFourLambdasII} show that 
the loading parameters $\lambda_{j1}=\frac{\lambda_{j1}}{1}=\frac{\lambda_{j1}}{\frac{\lambda_{11}+\hdots+\lambda_{41}}{4}}$ and 
$\lambda_{j2}=\frac{\lambda_{j2}}{1}=\frac{\lambda_{j2}}{\frac{\lambda_{12}+\hdots+\lambda_{42}}{4}}$ actually estimate the ratios
$\frac{A_1\to X_{j1}}{\frac{(A_1\to X_{11})+\hdots+(A_1\to X_{41})}{4}}$ and 
$\frac{A_2\to X_{j2}}{\frac{(A_2\to X_{12})+\hdots+(A_2\to X_{42})}{4}}$
($j=1,\hdots,4$). Introducing the notations $\overline{A_1\to X_{\bullet 1}}$ and $\overline{A_2\to X_{\bullet 2}}$ 
for the average loadings $\frac{(A_1\to X_{11})+\hdots+(A_1\to X_{41})}{4}$ and $\frac{(A_2\to X_{12})+\hdots+(A_2\to X_{42})}{4}$ corresponding
to $A_1$ and $A_2$, this can be rephrased as $\lambda_{j1}$ and $\lambda_{j2}$ estimating 
$\frac{A_1\to X_{j1}}{\overline{A_1\to X_{\bullet 1}}}$ and $\frac{A_2\to X_{j2}}{\overline{A_2\to X_{\bullet 2}}}$.
When effects coding is used for identification, estimated loading parameters 
hence describe by how much an indicator loads on a factor relative to 
how much that factor's indicators load on average. 

Combining Table~\ref{tab:IdentifiedFourCovariances} with Table~\ref{tab:ParametersFour} allows to infer which quantities
are actually estimated by $\Phi_{11},\Phi_{22},\Phi_{12}$, the parameters for latent variances (co-)variances: 
when the $i$-th indicator takes the role of the marker variable, they estimate $\operatorname{Var}(A_1)\cdot(A_1\to X_{i1})^2$, 
$\operatorname{Var}(A_2)\cdot(A_2\to X_{i2})^2$, and $\operatorname{Cov}(A_1,A_2)\cdot(A_1\to X_{i1})\cdot(A_2\to X_{i2})$, respectively.
Under the fixed factor method, $\Phi_{11}$ and $\Phi_{22}$ are fixed to unity, while
$\Phi_{12}$ estimates $\operatorname{Corr}(A_1,A_2)$, the correlation between $A_1$ and $A_2$.
Finally, when effects coding is used, the latent (co-)variance parameters measure
$\operatorname{Var}(A_1)\cdot\overline{A_1\to X_{\bullet 1}}^2$, 
$\operatorname{Var}(A_2)\cdot\overline{A_2\to X_{\bullet 2}}^2$, and 
$\operatorname{Cov}(A_1,A_2)\cdot\overline{A_1\to X_{\bullet 1}}\cdot\overline{A_2\to X_{\bullet 2}}$.

For the reader's convenience, all the interpretations given above are
summarized in Table~\ref{tab:InterFourI} in the appendix.


%% file: TableResults_TwoIndicators.tex
\begin{table}[ht]
\centering
\caption{Estimated model parameters for first example (unrestricted model), depending on alternative scaling methods.} 
\label{tab:ParametersTwo}
\begingroup\small
\begin{tabular}{lrrr}
  \toprule
 & Fixed Marker & Fixed Factor & Effects Coding \\ 
  \midrule
$\lambda_1$ & 1.00000 & 3.39411 & 1.23077 \\ 
  $\lambda_2$ & 0.62500 & 2.12132 & 0.76923 \\ 
  $\lambda_3$ & 1.00000 & 1.38564 & 1.23077 \\ 
  $\lambda_4$ & 0.62500 & 0.86603 & 0.76923 \\ 
   \midrule
$\Phi_{AA}$ & 11.52000 & 1.00000 & 7.60500 \\ 
  $\Phi_{BB}$ & 1.92000 & 1.00000 & 1.26750 \\ 
  $\Phi_{AB}$ & 3.20000 & 0.68041 & 2.11250 \\ 
   \midrule
$\Theta_{11}$ & 13.48000 & 13.48000 & 13.48000 \\ 
  $\Theta_{22}$ & 4.50000 & 4.50000 & 4.50000 \\ 
  $\Theta_{33}$ & 2.08000 & 2.08000 & 2.08000 \\ 
  $\Theta_{44}$ & 3.25000 & 3.25000 & 3.25000 \\ 
   \bottomrule
\end{tabular}
\endgroup
\end{table}

%% file: Explanation.tex
We first explain the reasons behind the constraint interaction in the first example
and then turn our attention to the second, more complex example.

\subsection{Explaining the first example}
Recall from above that we are interested in testing whether the population loadings $A\to X_2$ and $B\to X_4$ coincide 
and that this hypothesis is investigated by imposing the condition $\lambda_2=\lambda_4$ when estimating the so-called restricted model.
Building on the results from the previous section, we now know that $\lambda_2$ and $\lambda_4$ measure different quantities,
depending on the scaling method: for the fixed marker method, they estimate
$\frac{A\to X_2}{A\to X_1}$ and $\frac{B\to X_4}{B\to X_3}$, 
for the fixed factor method, they estimate $(A\to X_2)\cdot\sqrt{\operatorname{Var}(A)}$
and $(B\to X_4)\cdot\sqrt{\operatorname{Var}(B)}$, and for
effects coding, they measure $\frac{A\to X_2}{\frac{(A\to X_1)+(A\to X_2)}{2}}$ and 
$\frac{B\to X_4}{\frac{(B\to X_3)+(B\to X_4)}{2}}$.
By estimating the restricted model which enforces $\lambda_2=\lambda_4$, the null hypothesis actually tested thus depends on the scaling method!
Actually, we test
\begin{equation}
\label{eq:H0TwoMarker}
H_0: \frac{A\to X_2}{A\to X_1} \,\, = \,\, \frac{B\to X_4}{B\to X_3}\quad\mbox{ vs. } \quad
H_1: \frac{A\to X_2}{A\to X_1} \,\, \neq \,\, \frac{B\to X_4}{B\to X_3}
\end{equation}
in case of the fixed marker method, 
\begin{equation}
\begin{split}
\label{eq:H0TwoFactor}
H_0: & (A\to X_2)\cdot\sqrt{\operatorname{Var}(A)} \,\, = \,\, (B\to X_4)\cdot\sqrt{\operatorname{Var}(B)}\quad\mbox{ vs. } \quad\\
H_1: & (A\to X_2)\cdot\sqrt{\operatorname{Var}(A)} \,\, \neq \,\, (B\to X_4)\cdot\sqrt{\operatorname{Var}(B)}
\end{split}
\end{equation}
in case of the fixed factor method, and 
\begin{equation}
\label{eq:H0TwoEffects}
H_0: \dfrac{A\to X_2}{\frac{(A\to X_1)+(A\to X_2)}{2}} \,\, = \,\, \dfrac{B\to X_4}{\frac{(B\to X_3)+(B\to X_4)}{2}}\quad\mbox{ vs. } \quad
H_1: \dfrac{A\to X_2}{\frac{(A\to X_1)+(A\to X_2)}{2}} \,\, \neq \,\, \dfrac{B\to X_4}{\frac{(B\to X_3)+(B\to X_4)}{2}}
\end{equation}
in case of effects coding.

While it is easy to show that equations~\eqref{eq:H0TwoMarker} and \eqref{eq:H0TwoEffects} are 
equivalent\footnote{See \ref{sec:eqTwoMarkerEffect}. This equivalence of the results under the fixed marker method and effects coding
is due to the fact that the corresponding factors have only two indicators. For the more complex example with four indicators per factor, 
such an equivalence does not hold, see below.}, 
it is obvious that equation~\eqref{eq:H0TwoFactor} is not equivalent to the former two
equations, because equation~\eqref{eq:H0TwoFactor} is the only equation in which the standard
deviations of $A$ and $B$ appear. 
It is thus no coincidence that the results of testing $A\to X_2\stackrel{?}{=}B\to X_4$ are identical for
the fixed marker and effects coding methods, while the results for the fixed factor method are strongly different
from those: the first two methods test whether the equivalent equations~\eqref{eq:H0TwoMarker}
and \eqref{eq:H0TwoEffects} hold, while the latter tests equation~\eqref{eq:H0TwoFactor}.

The key to understanding constraint interaction is 
the fact that equations \eqref{eq:H0TwoMarker}, \eqref{eq:H0TwoFactor}, and \eqref{eq:H0TwoEffects} are
different implementations of equation~\eqref{eq:H0_Two}: 
when using the fixed marker method, \eqref{eq:H0TwoMarker} is tested, when using the fixed 
factor method, \eqref{eq:H0TwoFactor} is tested, and when using the effects coding method, 
\eqref{eq:H0TwoEffects} is tested. 

It is easy to see that the null hypotheses of the equivalent equations~\eqref{eq:H0TwoMarker} and \eqref{eq:H0TwoEffects} can be rewritten as
$\frac{A \rightarrow X_2}{B\rightarrow X_4}=\frac{A\rightarrow X_1}{B\rightarrow X_3}$, 
while the null hypothesis of equation~\eqref{eq:H0TwoFactor} can be rewritten as
$\frac{A \rightarrow X_2}{B\rightarrow X_4}=\frac{\sqrt{\operatorname{Var}(B)}}{\sqrt{\operatorname{Var}(A)}}$.
These two hypotheses are the more different, the more 
$\frac{A\rightarrow X_1}{B\rightarrow X_3}$ and 
$\frac{\sqrt{\operatorname{Var}(B)}}{\sqrt{\operatorname{Var}(A)}}$ are different, or equivalently, 
the more 
$\frac{\frac{A\rightarrow X_1}{B\rightarrow X_3}}{\frac{\sqrt{\operatorname{Var}(B)}}{\sqrt{\operatorname{Var}(A)}}}=
\frac{\left(A\rightarrow X_1\right)\cdot\sqrt{\operatorname{Var}(A)}}
{\left(B\rightarrow X_3\right)\cdot\sqrt{\operatorname{Var}(B)}}$ 
diverges from $1$. 
As a result, constraint interaction is the more likely to occur, the more 
$\frac{\left(A\rightarrow X_1\right)\cdot\sqrt{\operatorname{Var}(A)}}
{\left(B\rightarrow X_3\right)\cdot\sqrt{\operatorname{Var}(B)}}$ 
is different from $1$. 
The numerator of this expression, $(A\to X_1)\cdot\sqrt{\operatorname{Var}(A)}$, is the quantity that is estimated
by $\lambda_1$ under the fixed factor method, while the denominator, $(B\to X_3)\cdot\sqrt{\operatorname{Var}(B)}$,
is estimated by $\lambda_3$ in that case, see the previous section. From Table~\ref{tab:ParametersTwo},
we thus find that, in this example, 
$\frac{\left(A\rightarrow X_1\right)\cdot\sqrt{\operatorname{Var}(A)}}
{\left(B\rightarrow X_3\right)\cdot\sqrt{\operatorname{Var}(B)}}$ is estimated
by $\frac{3.39411}{1.38564}=2.44949\gg 1$, resulting in strong constraint interaction
due to quite different hypotheses being tested.

\subsection{Explaining the second example}
We now investigate the reasons behind constraint interaction in the more complex CFA 
model with a factor with four indicators measured twice in a longitudinal context.
Recall that the null hypothesis under consideration is $A_1\to X_{21}=A_2\to X_{22}$, which is investigated
by enforcing $\lambda_{21}=\lambda_{22}$ when estimating the restricted model. 
From the previous section, however, we now know that $\lambda_{21}$ and $\lambda_{22}$ measure different quantities,
depending on the scaling method: 
when the first indicators are used as marker variables, they estimate
$\frac{A_1\to X_{21}}{A_1\to X_{11}}$ and $\frac{A_2\to X_{22}}{A_2\to X_{12}}$, 
when the third indicators are used as marker variables, they estimate
$\frac{A_1\to X_{21}}{A_1\to X_{31}}$ and $\frac{A_2\to X_{22}}{A_2\to X_{32}}$, 
when the fourth indicators are used as marker variables, they estimate
$\frac{A_1\to X_{21}}{A_1\to X_{41}}$ and $\frac{A_2\to X_{22}}{A_2\to X_{42}}$, 
for the fixed factor method, they estimate $(A_1\to X_{21})\cdot\sqrt{\operatorname{Var}(A_1)}$ and $(A_2\to X_{22})\cdot\sqrt{\operatorname{Var}(A_2)}$, 
and for effects coding, they measure 
$\frac{A_1\to X_{21}}{\overline{A_1\to X_{\bullet 1}}}$ and $\frac{A_2\to X_{21}}{\overline{A_2\to X_{\bullet 2}}}$.
By estimating the restricted model which enforces $\lambda_{21}=\lambda_{22}$, the null hypothesis actually tested therefore
depends on the scaling method. More precisely, we test
\begin{equation}
\label{eq:H0FourMarkerI}
H_0: \frac{A_1\to X_{21}}{A_1\to X_{11}} \,\, = \,\, \frac{A_2\to X_{22}}{A_2\to X_{12}}\quad\mbox{ vs. } \quad
H_1: \frac{A_1\to X_{21}}{A_1\to X_{11}} \,\, \neq \,\, \frac{A_2\to X_{22}}{A_2\to X_{12}}
\end{equation}
when using the first indicators as marker variables, 
\begin{equation}
\label{eq:H0FourMarkerIII}
H_0: \frac{A_1\to X_{21}}{A_1\to X_{31}} \,\, = \,\, \frac{A_2\to X_{22}}{A_2\to X_{32}}\quad\mbox{ vs. } \quad
H_1: \frac{A_1\to X_{21}}{A_1\to X_{31}} \,\, \neq \,\, \frac{A_2\to X_{22}}{A_2\to X_{32}}
\end{equation}
when using the third indicators as marker variables, 
\begin{equation}
\label{eq:H0FourMarkerIV}
H_0: \frac{A_1\to X_{21}}{A_1\to X_{41}} \,\, = \,\, \frac{A_2\to X_{22}}{A_2\to X_{42}}\quad\mbox{ vs. } \quad
H_1: \frac{A_1\to X_{21}}{A_1\to X_{41}} \,\, \neq \,\, \frac{A_2\to X_{22}}{A_2\to X_{42}}
\end{equation}
when using the fourth indicators as marker variables, 
\begin{equation}
\begin{split}
\label{eq:H0FourFactor}
H_0: & (A_1\to X_{21})\cdot\sqrt{\operatorname{Var}(A_1)} \,\, = \,\, (A_2\to X_{22})\cdot\sqrt{\operatorname{Var}(A_2)}\quad\mbox{ vs. } \quad\\
H_1: & (A_1\to X_{21})\cdot\sqrt{\operatorname{Var}(A_1)} \,\, \neq \,\, (A_2\to X_{22})\cdot\sqrt{\operatorname{Var}(A_2)}
\end{split}
\end{equation}
in case of the fixed factor method, and 
\begin{equation}
\label{eq:H0FourEffects}
H_0: \dfrac{A_1\to X_{21}}{\overline{A_1\to X_{\bullet 1}}} \,\, = \,\, \dfrac{A_2\to X_{22}}{\overline{A_2\to X_{\bullet 2}}}\quad\mbox{ vs. } \quad
H_1: \dfrac{A_1\to X_{21}}{\overline{A_1\to X_{\bullet 1}}} \,\, \neq \,\, \dfrac{A_2\to X_{22}}{\overline{A_2\to X_{\bullet 2}}}
\end{equation}
in case of effects coding.

It is easy to see that the null hypotheses in equations~\eqref{eq:H0FourMarkerI}-\eqref{eq:H0FourFactor}
are equivalent to 
$\frac{A_1\to X_{21}}{A_2\to X_{22}}= \frac{A_1\to X_{11}}{A_2\to X_{12}}$, 
$\frac{A_1\to X_{21}}{A_2\to X_{22}}= \frac{A_1\to X_{31}}{A_2\to X_{32}}$, 
$\frac{A_1\to X_{21}}{A_2\to X_{22}}= \frac{A_1\to X_{41}}{A_2\to X_{42}}$, 
and $\frac{A_1\to X_{21}}{A_2\to X_{22}}=\frac{\sqrt{\operatorname{Var}(A_2)}}{\sqrt{\operatorname{Var}(A_1)}} $, respectively.
In \ref{sec:eqFourMarkerEffect}, we show that equation~\eqref{eq:H0FourEffects} is equivalent to
\begin{equation}
\begin{split}
\label{eq:H0FourEffectsII}
H_0: & \dfrac{A_1\to X_{21}}{A_2\to X_{22}} \,\, = \,\, \dfrac{(A_1\to X_{11})+(A_1\to X_{31})+(A_1\to X_{41})}{(A_2\to X_{12})+(A_2\to X_{32})+(A_2\to X_{42})}\quad\mbox{ vs. } \quad\\
H_1: & \dfrac{A_1\to X_{21}}{A_2\to X_{22}} \,\, \neq \,\, \dfrac{(A_1\to X_{11})+(A_1\to X_{31})+(A_1\to X_{41})}{(A_2\to X_{12})+(A_2\to X_{32})+(A_2\to X_{42})}.
\end{split}
\end{equation}
The five terms against which $\frac{A_1\to X_{21}}{A_2\to X_{22}}$ is compared, 
$\frac{A_1\to X_{11}}{A_2\to X_{12}}$, $\frac{A_1\to X_{31}}{A_2\to X_{32}}$, $\frac{A_1\to X_{41}}{A_2\to X_{42}}$, 
$\frac{\sqrt{\operatorname{Var}(A_2)}}{\sqrt{\operatorname{Var}(A_1)}}$, and $\frac{(A_1\to X_{11})+(A_1\to X_{31})+(A_1\to X_{41})}{(A_2\to X_{12})+(A_2\to X_{32})+(A_2\to X_{42})}$,
are all different, except for rare cases where some of these terms may happen to coincide.
As a consequence, the five null hypotheses tested under different scaling methods are typically all different, too,
leading to the five $\chi^2$-difference statistics in Table~\ref{tab:ResultsFour} deviating from each other. 
The constraint interaction discussed at the end of subsection~\ref{subsec:ExampleFour} can thus be explained as follows: 
the highly significant $\chi^2$-difference statistics obtained when using the first indicators as marker variables or when choosing the effects coding scaling
indicate that the data speaks extremely strongly against the actually tested null hypotheses in equations~\eqref{eq:H0FourMarkerI} and \eqref{eq:H0FourEffects}, respectively.
Furthermore, the data speaks rather strongly against the null hypothesis in equation~\eqref{eq:H0FourFactor}, the hypothesis related to the fixed factor scaling, 
while the data is more or less in accord with the null hypothesis in equation~\eqref{eq:H0FourMarkerIV}, the hypothesis related to using the fourth indicators as marker variables.
Finally, the data is perfectly in line with the null hypothesis in equation~\eqref{eq:H0FourMarkerIII}, the hypothesis actually being tested when choosing the third indicators as marker variables.

\auslassen { Otze fragt sich, ob man die nun Folgende in blau geschriebene verbale Erläuterung braucht?
{\color{blue}
the data speaks 
extremely strongly against hypothesis~\eqref{eq:H0FourMarkerI} with a $\chi^2$-difference statistic exceeding $145$ 
and a $p$-value of $0$, 
very strongly against hypothesis~\eqref{eq:H0FourEffects} with a $\chi^2$-difference statistic of above $26$
and a $p$-value of $0$, 
and rather strongly against hypothesis~\eqref{eq:H0FourFactor} with a $\chi^2$-difference statistic of about $5.58$
and a $p$-value of about $0.018$; 
the data is more or less in line with hypothesis~\eqref{eq:H0FourMarkerIV} with a $\chi^2$-difference statistic of about $2.72$
and a $p$-value of roughly $0.099$, and it is perfectly in line with hypothesis~\eqref{eq:H0FourMarkerIII}
with a $\chi^2$-difference statistic of $0$ and a $p$-value of $1$.}}

It is also possible to explain the amount by which the results vary according to the scaling methods employed:
while the null hypothesis in equation~\eqref{eq:H0FourMarkerI} is equivalent to testing the null hypothesis
$\frac{A_1\to X_{21}}{A_2\to X_{22}}=\frac{A_1\to X_{11}}{A_2\to X_{12}}$,
the null hypothesis in equation~\eqref{eq:H0FourMarkerIII} is equivalent to investigating the null hypothesis
$\frac{A_1\to X_{21}}{A_2\to X_{22}}=\frac{A_1\to X_{31}}{A_2\to X_{32}}=
\frac{A_1\to X_{31}}{A_1\to X_{11}}\cdot
\frac{A_2\to X_{12}}{A_2\to X_{32}}
\cdot
\frac{A_1\to X_{11}}{A_2\to X_{12}}$. The difference between these two hypotheses stems
from the term 
$\frac{A_1\to X_{31}}{A_1\to X_{11}}\cdot\frac{A_2\to {X_{12}}}{A_2\to X_{32}}$, 
which according to Tables~\ref{tab:IdentifiedFourLambdas} and~\ref{tab:IdentifiedFourLambdasII} 
is estimated by $4 \cdot 1.25=5\gg 1$, 
rendering hypotheses~\eqref{eq:H0FourMarkerI} and~\eqref{eq:H0FourMarkerIII} extremely different.
As a result, using the first indicators as marker variables leads to diametrically opposed results 
as compared to choosing the third indicators as marker variables.
In contrast, there is a rather small discrepancy between the results obtained when using the third and fourth indicators as marker variables, respectively,
which can be explained as follows: 
the difference between the corresponding hypotheses in equations~\eqref{eq:H0FourMarkerIII} and~\eqref{eq:H0FourMarkerIV} 
can be 
attributed to the term 
$\frac{A_1\to X_{31}}{A_1\to X_{41}}\cdot\frac{A_2\to X_{42}}{A_2\to X_{32}}$, 
which according to Tables~\ref{tab:IdentifiedFourLambdas} and~\ref{tab:IdentifiedFourLambdasII} 
is estimated by $1.6 \cdot 0.5=0.8\approx 1$. This leads to a detectable, but moderate difference
between hypotheses~\eqref{eq:H0FourMarkerIII} and~\eqref{eq:H0FourMarkerIV}.

\input{sec43}

%% file: sec43.tex
\subsection{Avoiding Constraint Interaction?}
The two examples discussed above have something peculiar in common.
In both cases, different scaling methods lead to different hypotheses actually being tested,
but none of the tested hypotheses is exactly equal or at least equivalent to the hypothesis that
should originally be tested: in fact, none of the three scaling methods in the first example leads to 
the original hypothesis~\eqref{eq:H0_Two} being tested, and none of the five different scaling methods
in the second example actually tests the original hypothesis~\eqref{eq:H0_Four}.
In view of this striking fact, the following question arises: 
is it possible to somehow cleverly design a scaling method whose use enables one to actually test the 
original hypotheses~\eqref{eq:H0_Two} and~\eqref{eq:H0_Four}?
Unfortunately, the answer to this question is negative: it is impossible to empirically test
the hypotheses~\eqref{eq:H0_Two} and~\eqref{eq:H0_Four}. 

In the following, we will explain in detail why it is impossible to empirically test 
whether the population loadings $A\to X_2$ and $B\to X_4$ coincide in the first example.\footnote{Completely analogous
reasoning shows why it is impossible to empirically test whether $A_1\to X_{21}$ and
$A_2\to X_{22}$ coincide in the second example.}
To this end, recall the fundamental principle underlying any statistical procedure for testing hypotheses: 
from observing sample data, one gains information about the distribution of the observed variables
and uses this information to make a decision between the null hypothesis and the alternative.
Such a decision is obviously only possible if the distribution of the data under the null 
hypothesis is different from the data's distribution under the alternative, as otherwise the data are
in no way informative with respect to telling apart the null and alternative hypotheses. 
For the example at hand, however, we show in \ref{sec:ModelImplied} that
the model-implied covariance matrix is identical for two 
different set of population quantities, of which one fulfills the
null hypothesis $A\to X_2=B\to X_4$, while for the other $A\to X_2\neq B\to X_4$.
As they share the model-implied covariance matrix, both these sets of population quantities imply the same distribution of
the manifest variables, making it impossible to tell these sets apart by 
observing the sample data. 
As a result, it is impossible to test the original hypothesis~\eqref{eq:H0_Two} based on observations
of the manifest variables $X_1,\hdots,X_4$. 


%% file: Conclusion.tex
By revisiting constraint interaction in the context of CFA models, this paper 
elaborates on the reasons underlying constraint interaction. 
In particular, constraint interaction is found to emerge both 
in conventional and longitudinal contexts as well as in examples with reasonable 
numbers of indicators per factor.
The reason behind constraint interaction is that 
different scaling methods lead to different hypotheses
being tested empirically. 
In order to find out which hypotheses are actually tested, 
it is essential to know how estimates of model parameters must be interpreted
depending on the method used for scaling the factors. 
While estimates of residual variances and covariances do not depend on the scaling method and 
always measure their population counterpart,
estimates of factor loadings and latent variances and covariances are sensitive
to the choice of scaling method, measuring different transformations of population quantities when
the scaling method is altered.
When the fixed marker method is used, the marker variable's loading in the population 
appears in some form in what other indicators' loading parameters estimate as well as in 
what is measured by the corresponding latent factor's estimated variances and covariances. 
Under the fixed factor method, the standard deviation of the latent factor affects estimated factor loadings, 
while the effects coding method is characterized by the population value of the latent factor's average loading
appearing both in estimated loadings and estimated latent variances and covariances.

In line with \citet{Steiger}, we recommend that researchers should be very careful about the scaling methods to be used and the interpretation
of estimated values of model parameters. 
When the fixed marker method is employed, an estimated loading parameter must be interpreted 
as an estimate of the ratio of the interesting loading over the corresponding marker variable's loading, 
while loading parameters estimated under the fixed factor method must be interpreted 
as estimates of the product of the loading of interest and the corresponding factor's standard deviation. 
Finally, when scaling is done via the effects coding method, an estimated loading parameter 
measures the ratio of the loading of interest to the average loading of the corresponding factor's indicators.

An important insight of this paper is that constraint interaction appears when trying to test
an hypothesis that 
can \emph{not} be tested empirically. In that case, empirical test procedures will not test the given hypothesis
as originally intended, 
but only a somehow related hypothesis whose exact form depends on the method used for scaling
the involved factor(s). Therefore, whenever constraint interaction appears in practice, 
this should be taken as a serious warning which indicates that one tries to test an hypothesis that 
can \emph{not} be tested by using the model at hand. 

In contrast, the phenomenon of constraint interaction may be helpful for detecting hypotheses 
that can not be tested empirically. 
When in doubt whether the hypothesis under consideration is 
empirically testable, researchers may conduct several tests of this hypothesis using different scaling methods:
if the results vary with the scaling method and constraint interaction is thus present, one can conclude
that the hypothesis under investigation is actually not empirically testable.

%% file: Appendix.tex
\subsection*{Equivalence of formulas~(\protect\ref{eq:H0TwoMarker}) and (\protect\ref{eq:H0TwoEffects})}
\label{sec:eqTwoMarkerEffect}
The following derivations show that the null hypothesis
$\frac{A\to X_2}{\frac{(A\to X_1)+(A\to X_2)}{2}} = \frac{B\to X_4}{\frac{(B\to X_3)+(B\to X_4)}{2}}$ 
is equivalent to the null hypothesis
$\frac{A\to X_2}{A\to X_1} = \frac{B\to X_4}{B\to X_3}$:
\begin{eqnarray*}
\dfrac{A\to X_2}{\frac{(A\to X_1)+(A\to X_2)}{2}} & = & \dfrac{B\to X_4}{\frac{(B\to X_3)+(B\to X_4)}{2}} \\[0.5em]
\Leftrightarrow  \frac{A\to X_2}{(A\to X_1)+(A\to X_2)} & = & \frac{B\to X_4}{(B\to X_3)+(B\to X_4)} \\[0.5em]
\Leftrightarrow  (A\to X_2) \cdot  \bigl( (B\to X_3)+(B\to X_4) \bigr) & = & (B\to X_4) \cdot \bigl( (A\to X_1)+(A\to X_2) \bigr) \\[0.5em]
\Leftrightarrow  (A\to X_2) \cdot (B\to X_3) & = & (A\to X_1)\cdot (B\to X_4) \\[0.5em]
\Leftrightarrow  \frac{A\to X_2}{A\to X_1} & = & \frac{B\to X_4}{B\to X_3}.
\end{eqnarray*}

\subsection*{Equivalence of formulas~(\protect\ref{eq:H0FourEffects}) and (\protect\ref{eq:H0FourEffectsII})}
\label{sec:eqFourMarkerEffect}
The following derivations show that the null hypothesis
$\frac{A_1\to X_{21}}{\overline{A_1\to X_{\bullet 1}}} = \frac{A_2\to X_{22}}{\overline{A_2\to X_{\bullet 2}}}$
is equivalent to the null hypothesis
$\frac{A_1\to X_{21}}{A_2\to X_{22}} = \frac{(A_1\to X_{11})+(A_1\to X_{31})+(A_1\to X_{41})}{(A_2\to X_{12})+(A_2\to X_{32})+(A_2\to X_{42})}$.
\begin{eqnarray*}
\dfrac{A_1\to X_{21}}{\overline{A_1\to X_{\bullet 1}}} & = & \dfrac{A_2\to X_{22}}{\overline{A_2\to X_{\bullet 2}}}\\[0.5em]
\Leftrightarrow 
\dfrac{A_1\to X_{21}}{\frac{(A_1\to X_{11})+\hdots+(A_1\to X_{41})}{4}} 
& = & 
\dfrac{A_2\to X_{22}}{\frac{(A_2\to X_{12})+\hdots+(A_2\to X_{42})}{4}} \\[0.5em]
\Leftrightarrow 
{\textstyle
\frac{A_1\to X_{21}}{(A_1\to X_{11})+\hdots+(A_1\to X_{41})} }
& = & 
{\textstyle
\frac{A_2\to X_{22}}{(A_2\to X_{12})+\hdots+(A_2\to X_{42})} }\\[0.5em]
%
%
\Leftrightarrow 
{\scriptstyle
(A_1\to X_{21})\cdot \bigl((A_2\to X_{12})+\hdots+(A_2\to X_{42})\bigr) }
& = & 
{\scriptstyle (A_2\to X_{22})\cdot\bigl((A_1\to X_{11})+\hdots+(A_1\to X_{41})\bigr) }\\[0.5em]
\Leftrightarrow 
{\scriptstyle
(A_1\to X_{21})\cdot \bigl((A_2\to X_{12})+(A_2\to X_{32})+(A_2\to X_{42})\bigr) }
& = & 
{\scriptstyle (A_2\to X_{22})\cdot\bigl((A_1\to X_{11})+(A_1\to X_{31})+(A_1\to X_{41})\bigr) }\\[0.5em]
\Leftrightarrow
\dfrac{A_1\to X_{21}}{A_2\to X_{22}} & = & 
\dfrac{(A_1\to X_{11})+(A_1\to X_{31})+(A_1\to X_{41})}{(A_2\to X_{12})+(A_2\to X_{32})+(A_2\to X_{42})}.
\end{eqnarray*}

\subsection*{Details of calculating the model-implied covariance matrix}
\label{sec:ModelImplied}
Let population loadings, latent (co-)variances, and residual variances be given by the entries of
Table~\ref{tab:ParametersTwo}'s first column, i.e.\ by
\begin{equation*}
\Lambda=
\begin{pmatrix}
A\to X_1 & 0\\
A\to X_2 & 0\\
0 & B\to X_3\\
0 & B\to X_4\\
\end{pmatrix}
=
\begin{pmatrix}
1 & 0\\
0.625 & 0\\
0 & 1\\
0 & 0.625\\
\end{pmatrix},
\end{equation*}

\begin{equation*}
\Phi=
\begin{pmatrix}
\operatorname{Var}(A) & \operatorname{Cov}(A,B)\\
\operatorname{Cov}(A,B) & \operatorname{Var}(B)\\
\end{pmatrix}
=
\begin{pmatrix}
11.52 & 3.2\\
3.2 & 1.92\\
\end{pmatrix},
\end{equation*}

\begin{equation*}
\Theta=
\begin{pmatrix}
\operatorname{Var}(E_1) & 0 & 0 & 0\\
0 & \operatorname{Var}(E_2) & 0 & 0\\
0 & 0 & \operatorname{Var}(E_3) & 0\\
0 & 0 & 0 & \operatorname{Var}(E_4)\\
\end{pmatrix}
=
\begin{pmatrix}
13.48 & 0 & 0 & 0\\
0 & 4.5 & 0 & 0\\
0 & 0 & 2.08 & 0\\
0 & 0 & 0 & 3.25\\
\end{pmatrix}.
\end{equation*}
Then the corresponding model-implied covariance matrix, $\Lambda\Phi\Lambda^{\prime}+\Theta$ (e.g., \citealp{Bollen1989}), coincides with the
matrix $S$ as given by equation~\eqref{eq:SigmaTwo}.

Alternatively, let the corresponding matrices be derived from the quantities of Table~\ref{tab:ParametersTwo}'s second
column, i.e.\ by
\begin{equation*}
\Lambda=
\begin{pmatrix}
A\to X_1 & 0\\
A\to X_2 & 0\\
0 & B\to X_3\\
0 & B\to X_4\\
\end{pmatrix}
=
\begin{pmatrix}
3.39411 & 0\\
2.12132 & 0\\
0 & 1.38564\\
0 & 0.86603\\
\end{pmatrix},
\end{equation*}

\begin{equation*}
\Phi=
\begin{pmatrix}
\operatorname{Var}(A) & \operatorname{Cov}(A,B)\\
\operatorname{Cov}(A,B) & \operatorname{Var}(B)\\
\end{pmatrix}
=
\begin{pmatrix}
1 & 0.68041\\
0.68041 & 1\\
\end{pmatrix},
\end{equation*}

\begin{equation*}
\Theta=
\begin{pmatrix}
\operatorname{Var}(E_1) & 0 & 0 & 0\\
0 & \operatorname{Var}(E_2) & 0 & 0\\
0 & 0 & \operatorname{Var}(E_3) & 0\\
0 & 0 & 0 & \operatorname{Var}(E_4)\\
\end{pmatrix}
=
\begin{pmatrix}
13.48 & 0 & 0 & 0\\
0 & 4.5 & 0 & 0\\
0 & 0 & 2.08 & 0\\
0 & 0 & 0 & 3.25\\
\end{pmatrix}.
\end{equation*}

Then -- up to rounding imprecision -- 
the corresponding model-implied covariance matrix, $\Lambda\Phi\Lambda^{\prime}+\Theta$, coincides with the
matrix $S$ as given by equation~\eqref{eq:SigmaTwo}.

%% file: AdditionalFigures.tex
\input{TableIdentified_TwoIndicators}

\input{TableResults_FourIndicators}
\input{TableIdentifiedLambdas_FourIndicators}
\input{TableIdentifiedLambdas_FourIndicatorsII}
\input{TableIdentifiedLambdas_FourCovariances}

\begin{landscape}
\input{TableInter_Two}
\end{landscape}

\begin{landscape}
\input{TableInter_FourI}
\end{landscape}

%% file: TableIdentified_TwoIndicators.tex
\begin{table}[h!t]
\centering
\caption{Estimated combinations of model parameters for first example (unrestricted model), depending on alternative scaling methods.} 
\label{tab:Identified}
\begingroup\small
\begin{tabular}{lrrr}
  \toprule
 & Fixed Marker & Fixed Factor & Effects Coding \\ 
  \midrule
$\dfrac{\lambda_2}{\lambda_1}$ & 0.62500 & 0.62500 & 0.62500 \\ 
   [1.1em]$\dfrac{\lambda_4}{\lambda_3}$ & 0.62500 & 0.62500 & 0.62500 \\ 
   [0.7em] \midrule
$\lambda_1\sqrt{\Phi_{AA}}$ & 3.39411 & 3.39411 & 3.39411 \\ 
   [0.5em]$\lambda_2\sqrt{\Phi_{AA}}$ & 2.12132 & 2.12132 & 2.12132 \\ 
   [0.5em]$\lambda_3\sqrt{\Phi_{BB}}$ & 1.38564 & 1.38564 & 1.38564 \\ 
   [0.5em]$\lambda_4\sqrt{\Phi_{BB}}$ & 0.86603 & 0.86603 & 0.86603 \\ 
   [0em] \midrule
$\dfrac{\lambda_1}{(\lambda_1+\lambda_2)/2}$ & 1.23077 & 1.23077 & 1.23077 \\ 
   [1.1em]$\dfrac{\lambda_2}{(\lambda_1+\lambda_2)/2}$ & 0.76923 & 0.76923 & 0.76923 \\ 
   [1.1em]$\dfrac{\lambda_3}{(\lambda_3+\lambda_4)/2}$ & 1.23077 & 1.23077 & 1.23077 \\ 
   [1.1em]$\dfrac{\lambda_4}{(\lambda_3+\lambda_4)/2}$ & 0.76923 & 0.76923 & 0.76923 \\ 
   [0.7em] \midrule
 \midrule
$\Phi_{AA}\lambda_1^2$ & 11.52000 & 11.52000 & 11.52000 \\ 
   [0.6em]$\Phi_{BB}\lambda_3^2$ & 1.92000 & 1.92000 & 1.92000 \\ 
   [0.6em]$\Phi_{AB}\lambda_1\lambda_3$ & 3.20000 & 3.20000 & 3.20000 \\ 
   [0em] \midrule
$\dfrac{\Phi_{AB}}{\sqrt{\Phi_{AA}\Phi_{BB}}}$ & 0.68041 & 0.68041 & 0.68041 \\ 
   [0.7em] \midrule
$\Phi_{AA}\cdot\left(\dfrac{\lambda_1+\lambda_2}{2}\right)^2$ & 7.60500 & 7.60500 & 7.60500 \\ 
   [1.1em]$\Phi_{BB}\cdot\left(\dfrac{\lambda_3+\lambda_4}{2}\right)^2$ & 1.26750 & 1.26750 & 1.26750 \\ 
   [1.1em]$\Phi_{AB}\cdot\dfrac{\lambda_1+\lambda_2}{2}\cdot\dfrac{\lambda_3+\lambda_4}{2}$ & 2.11250 & 2.11250 & 2.11250 \\ 
   [0.7em] \bottomrule
\end{tabular}
\endgroup
\end{table}

%% file: TableResults_FourIndicators.tex
\begin{table}[ht]
\centering
\caption{Estimated model parameters for second example (unrestricted model), depending on alternative scaling methods.} 
\label{tab:ParametersFour}
\begingroup\small
\begin{tabular}{lrrrrr}
  \toprule
 & Marker 1 & Marker 3 & Marker 4 & Factor & Effects \\ 
  \midrule
$\lambda_{11}$ & 1.000 & 0.250 & 0.400 & 0.800 & 0.320 \\ 
   [ 0.3 em]$\lambda_{21}$ & 5.000 & 1.250 & 2.000 & 4.000 & 1.600 \\ 
   [ 0.3 em]$\lambda_{31}$ & 4.000 & 1.000 & 1.600 & 3.200 & 1.280 \\ 
   [ 0.3 em]$\lambda_{41}$ & 2.500 & 0.625 & 1.000 & 2.000 & 0.800 \\ 
   [ 0.3 em]$\lambda_{12}$ & 1.000 & 1.250 & 2.500 & 5.000 & 1.250 \\ 
   [ 0.3 em]$\lambda_{22}$ & 1.000 & 1.250 & 2.500 & 5.000 & 1.250 \\ 
   [ 0.3 em]$\lambda_{32}$ & 0.800 & 1.000 & 2.000 & 4.000 & 1.000 \\ 
   [ 0.3 em]$\lambda_{42}$ & 0.400 & 0.500 & 1.000 & 2.000 & 0.500 \\ 
   [ 0.1 em] \midrule
$\Phi_{11}$ & 0.640 & 10.240 & 4.000 & 1.000 & 6.250 \\ 
   [ 0.3 em]$\Phi_{22}$ & 25.000 & 16.000 & 4.000 & 1.000 & 16.000 \\ 
   [ 0.3 em]$\Phi_{12}$ & 0.960 & 3.072 & 0.960 & 0.240 & 2.400 \\ 
   [ 0.2 em] \midrule
$\Theta_{11,11}$ & 3.000 & 3.000 & 3.000 & 3.000 & 3.000 \\ 
   [ 0.3 em]$\Theta_{21,21}$ & 1.000 & 1.000 & 1.000 & 1.000 & 1.000 \\ 
   [ 0.3 em]$\Theta_{31,31}$ & 4.000 & 4.000 & 4.000 & 4.000 & 4.000 \\ 
   [ 0.3 em]$\Theta_{41,41}$ & 2.000 & 2.000 & 2.000 & 2.000 & 2.000 \\ 
   [ 0.3 em]$\Theta_{12,12}$ & 2.000 & 2.000 & 2.000 & 2.000 & 2.000 \\ 
   [ 0.3 em]$\Theta_{22,22}$ & 7.000 & 7.000 & 7.000 & 7.000 & 7.000 \\ 
   [ 0.3 em]$\Theta_{32,32}$ & 1.000 & 1.000 & 1.000 & 1.000 & 1.000 \\ 
   [ 0.3 em]$\Theta_{42,42}$ & 8.000 & 8.000 & 8.000 & 8.000 & 8.000 \\ 
   [ 0.3 em]$\Theta_{11,12}$ & 0.200 & 0.200 & 0.200 & 0.200 & 0.200 \\ 
   [ 0.3 em]$\Theta_{21,22}$ & 0.500 & 0.500 & 0.500 & 0.500 & 0.500 \\ 
   [ 0.3 em]$\Theta_{31,32}$ & 0.250 & 0.250 & 0.250 & 0.250 & 0.250 \\ 
   [ 0.3 em]$\Theta_{41,42}$ & 0.500 & 0.500 & 0.500 & 0.500 & 0.500 \\ 
   [ 0.3 em] \bottomrule
\end{tabular}
\endgroup
\end{table}

%% file: TableIdentifiedLambdas_FourIndicators.tex
\begin{table}[ht]
\centering
\caption{Estimated combinations of model parameters for second example (unrestricted model), depending on alternative scaling methods.} 
\label{tab:IdentifiedFourLambdas}
\begingroup\small
\begin{tabular}{lrrrrr}
  \toprule
 & Marker 1 & Marker 3 & Marker 4 & Factor & Effects \\ 
  \midrule
$\dfrac{\lambda_{21}}{\lambda_{11}}$ & 5.000 & 5.000 & 5.000 & 5.000 & 5.000 \\ 
   [ 1.1 em]$\dfrac{\lambda_{31}}{\lambda_{11}}$ & 4.000 & 4.000 & 4.000 & 4.000 & 4.000 \\ 
   [ 1.1 em]$\dfrac{\lambda_{41}}{\lambda_{11}}$ & 2.500 & 2.500 & 2.500 & 2.500 & 2.500 \\ 
   [ 0.7 em] \midrule
$\dfrac{\lambda_{11}}{\lambda_{31}}$ & 0.250 & 0.250 & 0.250 & 0.250 & 0.250 \\ 
   [ 1.1 em]$\dfrac{\lambda_{21}}{\lambda_{31}}$ & 1.250 & 1.250 & 1.250 & 1.250 & 1.250 \\ 
   [ 1.1 em]$\dfrac{\lambda_{41}}{\lambda_{31}}$ & 0.625 & 0.625 & 0.625 & 0.625 & 0.625 \\ 
   [ 0.7 em] \midrule
$\dfrac{\lambda_{11}}{\lambda_{41}}$ & 0.400 & 0.400 & 0.400 & 0.400 & 0.400 \\ 
   [ 1.1 em]$\dfrac{\lambda_{21}}{\lambda_{41}}$ & 2.000 & 2.000 & 2.000 & 2.000 & 2.000 \\ 
   [ 1.1 em]$\dfrac{\lambda_{31}}{\lambda_{41}}$ & 1.600 & 1.600 & 1.600 & 1.600 & 1.600 \\ 
   [ 0.7 em] \midrule
$\lambda_{11}\sqrt{\Phi_{11}}$ & 0.800 & 0.800 & 0.800 & 0.800 & 0.800 \\ 
   [ 0.5 em]$\lambda_{21}\sqrt{\Phi_{11}}$ & 4.000 & 4.000 & 4.000 & 4.000 & 4.000 \\ 
   [ 0.5 em]$\lambda_{31}\sqrt{\Phi_{11}}$ & 3.200 & 3.200 & 3.200 & 3.200 & 3.200 \\ 
   [ 0.5 em]$\lambda_{41}\sqrt{\Phi_{11}}$ & 2.000 & 2.000 & 2.000 & 2.000 & 2.000 \\ 
   [ 0 em] \midrule
$\dfrac{\lambda_{11}}{\left(\lambda_{11}+\hdots+\lambda_{41}\right)/4}$ & 0.320 & 0.320 & 0.320 & 0.320 & 0.320 \\ 
   [ 1.1 em]$\dfrac{\lambda_{21}}{\left(\lambda_{11}+\hdots+\lambda_{41}\right)/4}$ & 1.600 & 1.600 & 1.600 & 1.600 & 1.600 \\ 
   [ 1.1 em]$\dfrac{\lambda_{31}}{\left(\lambda_{11}+\hdots+\lambda_{41}\right)/4}$ & 1.280 & 1.280 & 1.280 & 1.280 & 1.280 \\ 
   [ 1.1 em]$\dfrac{\lambda_{41}}{\left(\lambda_{11}+\hdots+\lambda_{41}\right)/4}$ & 0.800 & 0.800 & 0.800 & 0.800 & 0.800 \\ 
   [ 0.7 em] \bottomrule
\end{tabular}
\endgroup
\end{table}

%% file: TableIdentifiedLambdas_FourIndicatorsII.tex
\begin{table}[ht]
\centering
\caption{Estimated combinations of model parameters for second example (unrestricted model), depending on alternative scaling methods.} 
\label{tab:IdentifiedFourLambdasII}
\begingroup\small
\begin{tabular}{lrrrrr}
  \toprule
 & Marker 1 & Marker 3 & Marker 4 & Factor & Effects \\ 
  \midrule
$\dfrac{\lambda_{22}}{\lambda_{12}}$ & 1.000 & 1.000 & 1.000 & 1.000 & 1.000 \\ 
   [ 1.1 em]$\dfrac{\lambda_{32}}{\lambda_{12}}$ & 0.800 & 0.800 & 0.800 & 0.800 & 0.800 \\ 
   [ 1.1 em]$\dfrac{\lambda_{42}}{\lambda_{12}}$ & 0.400 & 0.400 & 0.400 & 0.400 & 0.400 \\ 
   [ 0.7 em] \midrule
$\dfrac{\lambda_{12}}{\lambda_{32}}$ & 1.250 & 1.250 & 1.250 & 1.250 & 1.250 \\ 
   [ 1.1 em]$\dfrac{\lambda_{22}}{\lambda_{32}}$ & 1.250 & 1.250 & 1.250 & 1.250 & 1.250 \\ 
   [ 1.1 em]$\dfrac{\lambda_{42}}{\lambda_{32}}$ & 0.500 & 0.500 & 0.500 & 0.500 & 0.500 \\ 
   [ 0.7 em] \midrule
$\dfrac{\lambda_{12}}{\lambda_{42}}$ & 2.500 & 2.500 & 2.500 & 2.500 & 2.500 \\ 
   [ 1.1 em]$\dfrac{\lambda_{22}}{\lambda_{42}}$ & 2.500 & 2.500 & 2.500 & 2.500 & 2.500 \\ 
   [ 1.1 em]$\dfrac{\lambda_{32}}{\lambda_{42}}$ & 2.000 & 2.000 & 2.000 & 2.000 & 2.000 \\ 
   [ 0.7 em] \midrule
$\lambda_{12}\sqrt{\Phi_{22}}$ & 5.000 & 5.000 & 5.000 & 5.000 & 5.000 \\ 
   [ 0.5 em]$\lambda_{22}\sqrt{\Phi_{22}}$ & 5.000 & 5.000 & 5.000 & 5.000 & 5.000 \\ 
   [ 0.5 em]$\lambda_{32}\sqrt{\Phi_{22}}$ & 4.000 & 4.000 & 4.000 & 4.000 & 4.000 \\ 
   [ 0.5 em]$\lambda_{42}\sqrt{\Phi_{22}}$ & 2.000 & 2.000 & 2.000 & 2.000 & 2.000 \\ 
   [ 0 em] \midrule
$\dfrac{\lambda_{12}}{\left(\lambda_{12}+\hdots+\lambda_{42}\right)/4}$ & 1.250 & 1.250 & 1.250 & 1.250 & 1.250 \\ 
   [ 1.1 em]$\dfrac{\lambda_{22}}{\left(\lambda_{12}+\hdots+\lambda_{42}\right)/4}$ & 1.250 & 1.250 & 1.250 & 1.250 & 1.250 \\ 
   [ 1.1 em]$\dfrac{\lambda_{32}}{\left(\lambda_{12}+\hdots+\lambda_{42}\right)/4}$ & 1.000 & 1.000 & 1.000 & 1.000 & 1.000 \\ 
   [ 1.1 em]$\dfrac{\lambda_{42}}{\left(\lambda_{12}+\hdots+\lambda_{42}\right)/4}$ & 0.500 & 0.500 & 0.500 & 0.500 & 0.500 \\ 
   [ 0.7 em] \bottomrule
\end{tabular}
\endgroup
\end{table}

%% file: TableIdentifiedLambdas_FourCovariances.tex
\begin{table}[ht]
\centering
\caption{Estimated combinations of model parameters for second example (unrestricted model), depending on alternative scaling methods.} 
\label{tab:IdentifiedFourCovariances}
\begingroup\small
\begin{tabular}{lrrrrr}
  \toprule
 & Marker 1 & Marker 3 & Marker 4 & Factor & Effects \\ 
  \midrule
$\Phi_{11}\lambda_{11}^2$ & 0.640 & 0.640 & 0.640 & 0.640 & 0.640 \\ 
   [ 0.6 em]$\Phi_{22}\lambda_{12}^2$ & 25.000 & 25.000 & 25.000 & 25.000 & 25.000 \\ 
   [ 0.6 em]$\Phi_{12}\lambda_{11}\lambda_{12}$ & 0.960 & 0.960 & 0.960 & 0.960 & 0.960 \\ 
   [ 0.1 em] \midrule
$\Phi_{11}\lambda_{31}^2$ & 10.240 & 10.240 & 10.240 & 10.240 & 10.240 \\ 
   [ 0.6 em]$\Phi_{22}\lambda_{32}^2$ & 16.000 & 16.000 & 16.000 & 16.000 & 16.000 \\ 
   [ 0.6 em]$\Phi_{12}\lambda_{31}\lambda_{32}$ & 3.072 & 3.072 & 3.072 & 3.072 & 3.072 \\ 
   [ 0.1 em] \midrule
$\Phi_{11}\lambda_{41}^2$ & 4.000 & 4.000 & 4.000 & 4.000 & 4.000 \\ 
   [ 0.6 em]$\Phi_{22}\lambda_{42}^2$ & 4.000 & 4.000 & 4.000 & 4.000 & 4.000 \\ 
   [ 0.6 em]$\Phi_{12}\lambda_{41}\lambda_{42}$ & 0.960 & 0.960 & 0.960 & 0.960 & 0.960 \\ 
   [ 0.1 em] \midrule
$\dfrac{\Phi_{12}}{\sqrt{\Phi_{11}\Phi_{22}}}$ & 0.240 & 0.240 & 0.240 & 0.240 & 0.240 \\ 
   [ 0.1 em] \midrule
$\Phi_{11}\cdot\left(\dfrac{\lambda_{11}+\hdots+\lambda_{41}}{4}\right)^2$ & 6.250 & 6.250 & 6.250 & 6.250 & 6.250 \\ 
   [ 1.3 em]$\Phi_{22}\cdot\left(\dfrac{\lambda_{12}+\hdots+\lambda_{42}}{4}\right)^2$ & 16.000 & 16.000 & 16.000 & 16.000 & 16.000 \\ 
   [ 1.4 em]$\Phi_{12}\cdot\dfrac{\lambda_{11}+\hdots+\lambda_{41}}{4}\cdot\dfrac{\lambda_{12}+\hdots+\lambda_{42}}{4}$ & 2.400 & 2.400 & 2.400 & 2.400 & 2.400 \\ 
   [ 0.1 em] \bottomrule
\end{tabular}
\endgroup
\end{table}

%% file: TableInter_Two.tex
\begin{table}[ht]
\centering
\caption{Transformations of population quantities that model parameters estimate under alternative scaling methods in first example. $E_{i}$ ($i=1,\hdots,4$) denote the error terms of the indicators.} 
\label{tab:InterTwo}
\begingroup\small
\begin{tabular}{lccc}
  \toprule
 & Fixed Marker & Fixed Factor & Effects Coding \\ 
  \midrule
$\lambda_1$ & $\dfrac{A\rightarrow X_1}{A\rightarrow X_1}=1$ & $\left(A\rightarrow X_1\right)\cdot\sqrt{\operatorname{Var}(A)}$ & $\dfrac{A\rightarrow X_1}{\frac{\left(A\rightarrow X_1\right)+\left(A\rightarrow X_2\right)}{2}}$ \\ 
   [1.5em]$\lambda_2$ & $\dfrac{A\rightarrow X_2}{A\rightarrow X_1}$ & $\left(A\rightarrow X_2\right)\cdot\sqrt{\operatorname{Var}(A)}$ & $\dfrac{A\rightarrow X_2}{\frac{\left(A\rightarrow X_1\right)+\left(A\rightarrow X_2\right)}{2}}$ \\ 
   [1.5em]$\lambda_3$ & $\dfrac{B\rightarrow X_3}{B\rightarrow X_3}=1$ & $\left(B\rightarrow X_3\right)\cdot\sqrt{\operatorname{Var}(B)}$ & $\dfrac{B\rightarrow X_3}{\frac{\left(B\rightarrow X_3\right)+\left(B\rightarrow X_4\right)}{2}}$ \\ 
   [1.5em]$\lambda_4$ & $\dfrac{B\rightarrow X_4}{B\rightarrow X_3}$ & $\left(B\rightarrow X_4\right)\cdot\sqrt{\operatorname{Var}(B)}$ & $\dfrac{B\rightarrow X_4}{\frac{\left(B\rightarrow X_3\right)+\left(B\rightarrow X_4\right)}{2}}$ \\ 
   [1.5em] \midrule
$\Phi_{AA}$ & $\operatorname{Var}(A)\cdot\left(A\rightarrow X_1\right)^2$ & $\dfrac{\operatorname{Var}(A)}{\operatorname{Var}(A)}=1$ & $\operatorname{Var}(A)\cdot\left(\dfrac{\left(A\rightarrow X_1\right)+\left(A\rightarrow X_2\right)}{2}\right)^2$ \\ 
   [1.5em]$\Phi_{BB}$ & $\operatorname{Var}(B)\cdot\left(B\rightarrow X_3\right)^2$ & $\dfrac{\operatorname{Var}(B)}{\operatorname{Var}(B)}=1$ & $\operatorname{Var}(B)\cdot\left(\dfrac{\left(B\rightarrow X_3\right)+\left(B\rightarrow X_4\right)}{2}\right)^2$ \\ 
   [1.5em]$\Phi_{AB}$ & $\operatorname{Cov}(A,B)\cdot\left(A\rightarrow X_1\right)\cdot\left(B\rightarrow X_3\right)$ & $\operatorname{Corr}(A,B)$ & $\operatorname{Cov}(A,B)\cdot\dfrac{\left(A\rightarrow X_1\right)+\left(A\rightarrow X_2\right)}{2}\cdot\dfrac{\left(B\rightarrow X_3\right)+\left(B\rightarrow X_4\right)}{2}$ \\ 
   [1.5em] \midrule
$\Theta_{11}$ & $\operatorname{Var}(E_{1})$ & $\operatorname{Var}(E_{1})$ & $\operatorname{Var}(E_{1})$ \\ 
   [.5em]$\Theta_{22}$ & $\operatorname{Var}(E_{2})$ & $\operatorname{Var}(E_{2})$ & $\operatorname{Var}(E_{2})$ \\ 
   [.5em]$\Theta_{33}$ & $\operatorname{Var}(E_{3})$ & $\operatorname{Var}(E_{3})$ & $\operatorname{Var}(E_{3})$ \\ 
   [.5em]$\Theta_{44}$ & $\operatorname{Var}(E_{4})$ & $\operatorname{Var}(E_{4})$ & $\operatorname{Var}(E_{4})$ \\ 
   [.5em] \bottomrule
\end{tabular}
\endgroup
\end{table}

%% file: TableInter_FourI.tex
\begin{table}[ht]
\centering
\caption{Transformations of population quantities that model parameters estimate under alternative scaling methods in second example. \\ $E_{j1}, E_{j2}$ ($j=1,\hdots,4$) 
              denote the error terms of the indicators at times $1$ and $2$.} 
\label{tab:InterFourI}
\begingroup\small
\begin{tabular}{lccc}
  \toprule
 & Marker $i$ ($i=1,3,4$) & Factor & Effects \\ 
  \midrule
$\lambda_{j1}$ & $\dfrac{A_1\rightarrow X_{j1}}{A_1\rightarrow X_{i1}}$ & $\left(A_1\rightarrow X_{j1}\right)\cdot\sqrt{\operatorname{Var}(A_1)}$ & $\dfrac{A_1\rightarrow X_{j1}}{\overline{A_1\rightarrow X_{\bullet 1}}}$ \\ 
   [1.5em]$\lambda_{j2}$ & $\dfrac{A_2\rightarrow X_{j2}}{A_2\rightarrow X_{i2}}$ & $\left(A_2\rightarrow X_{j2}\right)\cdot\sqrt{\operatorname{Var}(A_2)}$ & $\dfrac{A_2\rightarrow X_{j2}}{\overline{A_2\rightarrow X_{\bullet 2}}}$ \\ 
   [1.5em] \midrule
$\Phi_{11}$ & $\operatorname{Var}(A_1)\cdot\left(A_1\rightarrow X_{i1}\right)^2$ & $\dfrac{\operatorname{Var}(A_1)}{\operatorname{Var}(A_1)}=1$ & $\operatorname{Var}(A_1)\cdot\overline{A_1\rightarrow X_{\bullet 1}}^2$ \\ 
   [1.5em]$\Phi_{22}$ & $\operatorname{Var}(A_2)\cdot\left(A_2\rightarrow X_{i2}\right)^2$ & $\dfrac{\operatorname{Var}(A_2)}{\operatorname{Var}(A_2)}=1$ & $\operatorname{Var}(A_2)\cdot\overline{A_2\rightarrow X_{\bullet 2}}^2$ \\ 
   [1.5em]$\Phi_{12}$ & $\operatorname{Cov}(A_1,A_2)\cdot\left( A_1\rightarrow X_{i1}\right)\cdot\left(A_2\rightarrow X_{i2}\right)$ & $\operatorname{Corr}(A_1,A_2)$ & $\operatorname{Cov}(A_1,A_2)\cdot\overline{A_1\rightarrow X_{\bullet 1}}\cdot\overline{A_2\rightarrow X_{\bullet2}}$ \\ 
   [1.5em] \midrule
$\Theta_{j1,j1}$ & $\operatorname{Var}(E_{j1})$ & $\operatorname{Var}(E_{j1})$ & $\operatorname{Var}(E_{j1})$ \\ 
   [.5em]$\Theta_{j2,j2}$ & $\operatorname{Var}(E_{j2})$ & $\operatorname{Var}(E_{j2})$ & $\operatorname{Var}(E_{j2})$ \\ 
   [.5em]$\Theta_{j1,j2}$ & $\operatorname{Cov}(E_{j1},E_{j2})$ & $\operatorname{Cov}(E_{j1},E_{j2})$ & $\operatorname{Cov}(E_{j1},E_{j2})$ \\ 
   [.5em] \bottomrule
\end{tabular}
\endgroup
\end{table}